\documentclass[aps, prl, reprint, superscriptaddress, nofootinbib]{revtex4-1}

\usepackage[utf8]{inputenc}
\usepackage[english]{babel}

\usepackage{hyperref}
\hypersetup{
	colorlinks=true,
	linkcolor=blue,
	urlcolor=red,
	citecolor=blue
}

\usepackage{lmodern}
\usepackage{graphicx}
\usepackage{amsmath}
\usepackage{amssymb}
\usepackage{upgreek}
\usepackage{mathrsfs}
\usepackage{verbatim}   % useful for program listings
\usepackage{subfigure}  % use for side-by-side figures
\usepackage{xcolor}
\usepackage{placeins}
\usepackage{color}

\usepackage{footmisc}

\begin{document}
	
%%%%%%%%%%%%%%%%%%
% Personal Stuff %
%%%%%%%%%%%%%%%%%%
\newcommand{\TEMOO}{TEM$_{00}~$}
\newcommand{\TEMIO}{TEM$_{10}~$}
\newcommand{\SiN}{Si$_3$N$_4~$}\
\newcommand{\opt}{_{\text{opt}}}
\newcommand{\eff}{_{\text{eff}}}
\newcommand{\tot}{_{\text{tot}}}
\newcommand{\mat}{_{\text{mat}}}
\newcommand{\unit}{_{\text{unit}}}
\newcommand{\others}[1]{}
\newcommand{\?}[1]{\textcolor{red}{[#1]}}

\title{Probing a spin transfer controlled magnetic nanowire with a single nitrogen-vacancy spin in bulk diamond}

\author{Adrian Solyom\footref{fn}}
\author{Zackary Flansberry\footref{fn}}
\author{M\"{a}rta A. Tschudin}
\author{Nathaniel Leitao}
\affiliation{McGill University Department of Physics}

% Define new footnote style to set author contributions
\renewcommand{\thefootnote}{\fnsymbol{footnote}}
\footnotetext[2]{These authors contributed equally. \label{fn}}
\footnotetext[1]{lilian.childress@mcgill.ca \label{corresponding-lily}}

\author{Michel Pioro-Ladri\`{e}re}
\affiliation{Universit\'{e} de Sherbrooke Institut Quantique and D\'{e}partement de Physique}
\affiliation{Quantum Information Science Program, Canadian Institute for Advanced Research}

\author{Jack C. Sankey}
\author{Lilian I. Childress\footref{corresponding-lily}}

\affiliation{McGill University Department of Physics}

\date{\today}

\begin{abstract}
The point-like nature and exquisite magnetic field sensitivity of the nitrogen vacancy  (NV) center in diamond can provide information about the inner workings of magnetic nanocircuits in complement with traditional transport techniques. Here we use a single NV in bulk diamond to probe the stray field of a ferromagnetic nanowire controlled by spin transfer (ST) torques. We first report an unambiguous measurement of ST tuned, parametrically driven, large-amplitude magnetic oscillations. At the same time, we demonstrate that such magnetic oscillations alone can directly drive NV spin transitions, providing a potential new means of control. Finally, we use the NV as a local noise thermometer, observing strong ST damping of the stray field noise, consistent with magnetic cooling from room temperature to $\sim$150~K.
\end{abstract}

\maketitle

%%%%%%%%%%%%%%%%%%%%%%%%%%%%%%%%%%%%%%%%%%%
% INTRODUCTION
%%%%%%%%%%%%%%%%%%%%%%%%%%%%%%%%%%%%%%%%%%%

In nanoscale magnetic circuits, spin transfer (ST) effects provide an efficient means of all-electronic control, prompting the development of ST-based non-volatile memories, microwave-frequency oscillators, filters, detectors, and amplifiers \cite{Ralph2008Spin, Manchon2018CurrentARXIV}. These developments call for new tools to understand the interplay between spins, magnons, and the environment. In parallel, the nitrogen-vacancy (NV) center in diamond~\cite{Doherty2013The} has emerged as a versatile sensor for studying magnetic systems, with excellent spatial and spectral resolution~\cite{Rondin2014Magnetometry, Casola2018Probing}. The tiny magnetic moment of a single spin provides a non-invasive probe of local stray fields, and can be combined with scanned-probe~\cite{Maletinsky2012A} or subwavelength imaging~\cite{Rittweger2009STED} techniques to achieve nanoscale spatial resolution. Moreover, the NV offers a variety of sensing modalities appropriate for DC and AC magnetic fields~\cite{Degen2008Scanning, Taylor2008High} or noise spectroscopy~\cite{Tetienne2013Spin, Rosskopf2014Investigation, Myers2014Probing}. Recently, NV centers have probed ferromagnetic phenomena including vortex cores~\cite{Tetienne2013Quantitative, Badea2016Exploiting}, domain walls~\cite{Tetienne2014Nanoscale},  ferromagnetic resonance (FMR)~\cite{Wolfe2014Off, Van2015Nanometre, Wolfe2016Spatially, Page2016OpticallyARXIV, Du2017Control}, and magnetic thermal quantities in an extended YIG film \cite{Du2017Control} modified by ST effects.

Here, we probe a metallic ST-controlled ferromagnetic nanowire with a single NV in bulk diamond. Different from microwave-frequency magnetoresistance readout \cite{Kiselev2003Microwave, Tulapurkar2005Spin, Sankey2006Spin}, the NV's point-like nature enables coupling to shorter-wavelength spin waves, thereby revealing qualitatively new features. First, we provide transport-based evidence of parametrically-driven magnetic oscillations -- tuned through threshold with spin transfer anti-damping -- and use the resulting (phase-locked) stray field oscillations \emph{alone} to drive the NV spin resonance. 
\begin{figure}[h]
	\includegraphics[width=0.99\columnwidth]{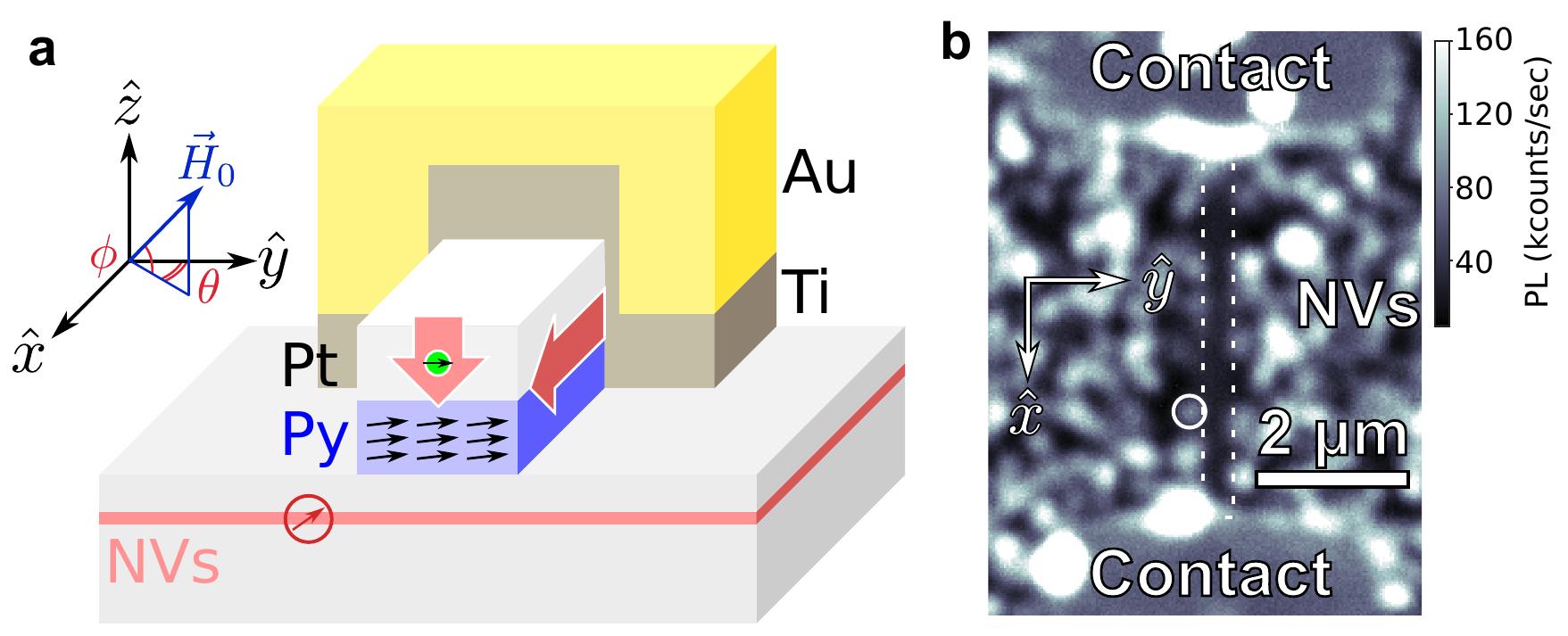}
	\caption{Device geometry. (a) Cross-section of nanowire on a diamond substrate with implanted NVs. Nanowire current (red arrow along $\hat{x}$) drives spins (green circle) polarized along $\hat{y}$ into the Py layer (red arrow along $\hat{z}$). (b) Confocal PL map of the device.	The nanowire (dotted box) appears as a shadow. Circle indicates the NV under study.}
	\label{fig1}
\end{figure}
\begin{figure*}
	\includegraphics[width=0.95\textwidth]{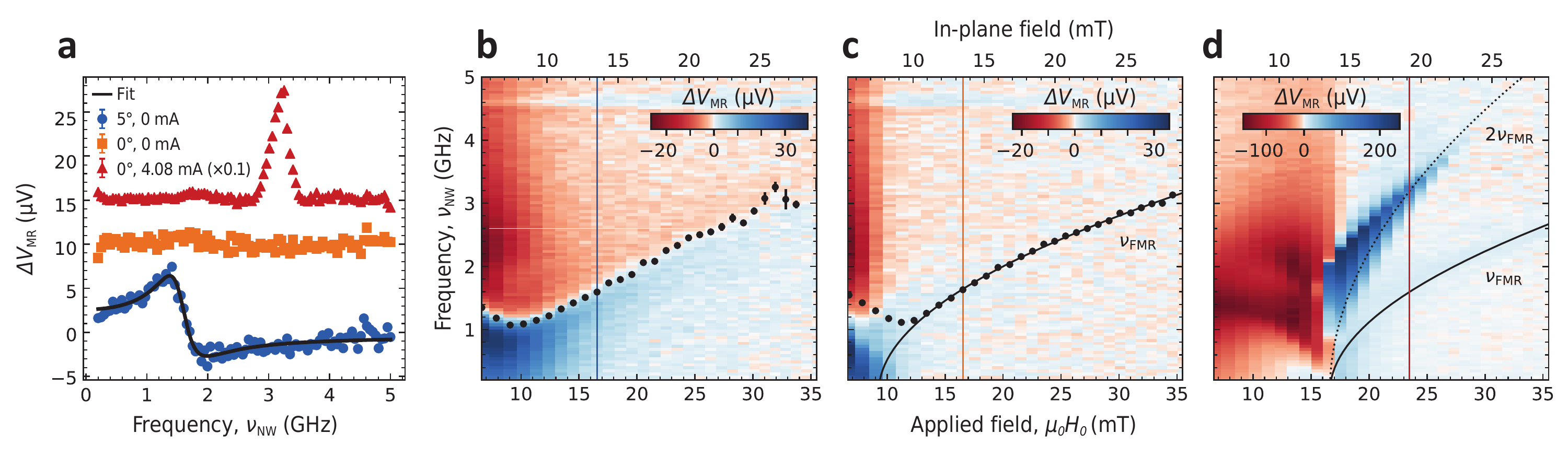}
	\caption{Transport FMR and parametric response. (a) Voltage $\Delta V_\text{MR}$ spectra (vertically separated for clarity) for bias $I_0=0$, RF amplitude $I_\text{RF}=0.95$~mA, and  magnetic field $\mu_0 H_0=16.5$~mT applied along the NV axis (orange) or rotated about $\hat{z}$ by $\theta=5^\circ$ (blue). Red points ($y$-values divided by 10) correspond to $I_0=4.08$~mA and $\theta=0^\circ$ (taken at  $\mu_0H_0=23.5$~mT), showing a large resonance near twice the FMR frequency $\nu_\text{FMR}$. (b) and (c) show the frequency and field dependence of $\Delta V_\text{MR}$ for $\theta=5^\circ$ and $0^\circ$, with symbols denoting fit values of $\nu_\text{FMR}$. (d) Same as (c), but with $I_0= 4.08$~mA. Vertical lines in (b)-(d) correspond to the traces plotted in (a).}
	\label{fig2}
\end{figure*}
This demonstration removes any ambiguity from the interpretation of the transport measurements, and provides a potential new means of NV control. Additionally, transport readout indicates a precession angle up to $\sim$55$^\circ$, suggesting the absence of Suhl-like (multi-magnon) instabilities \cite{Suhl1957The}, which we tentatively attribute to the reduced density of states in our confined geometry. The observed changes in field at the NV independently corroborate the estimated angle, and (more importantly) we observe no signs of NV spin flips in the PL spectra when directly driving FMR -- a feature normally appearing in extended magnetic films \cite{Wolfe2014Off, Van2015Nanometre, Wolfe2016Spatially, Page2016OpticallyARXIV, Du2017Control} -- providing additional evidence of suppressed multi-magnon instabilities. This highlights the NV's potential use as a tool for understanding internal workings of magnetic nanosystems. Finally, we demonstrate strong ST control of the magnetic thermal fluctuations: adapting the NV noise relaxometry method of Ref.~\cite{Du2017Control}, we observe large ST damping of the stray field noise, providing evidence of magnetic cooling from room temperature to $\sim$150~K. The observed noise suppression is orders of magnitude larger than measured with an NV near a YIG film \cite{Du2017Control}, and a factor of $\sim$2 larger than measured with Brillouin light scattering from a related metallic structure \cite{Demidov2017Chemical}. An interesting open question is the fundamental limits of this cooling technique, in particular if and from what temperature the ground state can be reached. Interestingly, our demonstrated NV sensitivity suggests it may be possible to resolve magnetic zero-point fluctuations, thereby providing an optically active ``handle'' on a macroscopic magnet in the quantum regime.

These techniques, which are especially advantageous in systems having low magnetoresistance, pave the way to a deeper understanding of short-wavelength magnons, ST effects, and quantum emitter control.

%%%%%%%%%%%%%%%%%%%%%%%%%%%%%%%%%%%%%%%%%%%%%%%
% GEOMETRY OF SYSTEM, BASICS
%%%%%%%%%%%%%%%%%%%%%%%%%%%%%%%%%%%%%%%%%%%%%%%

\section{Device Geometry}

We fabricate \cite{supplementary} an $8$-$\upmu$m-long $\times$ $417$-nm-wide Py ($\mathrm{Ni_{81}Fe_{19}}$, 10 nm) / Pt (10 nm) multilayer nanowire on electronic grade diamond with a layer of NVs implanted $60\pm 15$~nm below the surface, as shown in Fig.~\ref{fig1}(a). Figure \ref{fig1}(b) shows a photoluminescence (PL) image of the device pumped (532~nm) and collected ($>$594~nm) from above. Bright spots are NVs, the nanowire appears dark, and the contacts (Au) exhibit typical background PL. We focus on the indicated spot, with PL primarily from single NV whose symmetry axis lies $35^\circ$ from the $\hat{y}$-axis in the $yz$ plane \cite{supplementary}. Static magnetic fields are applied with a permanent neodymium magnet on a motorized stage, calibrated using the NV spin transitions \cite{supplementary}. For this device, a moderate field ($10-40$~mT) applied along the NV symmetry axis saturates the Py magnetization along $\hat{y}$, canted at most $\sim$3$^\circ$ out of the $xy$ plane (dictated by shape anisotropy \cite{supplementary}). All measurements are taken at room temperature. 

%%%%%%%%%%%%%%%%%%%%%%%%%%%%%%%%%%%%%%%%%
% TRANSPORT FMR AND PARAMETRIC
%%%%%%%%%%%%%%%%%%%%%%%%%%%%%%%%%%%%%%%%%

\begin{figure*}
	\includegraphics[width=0.95\textwidth]{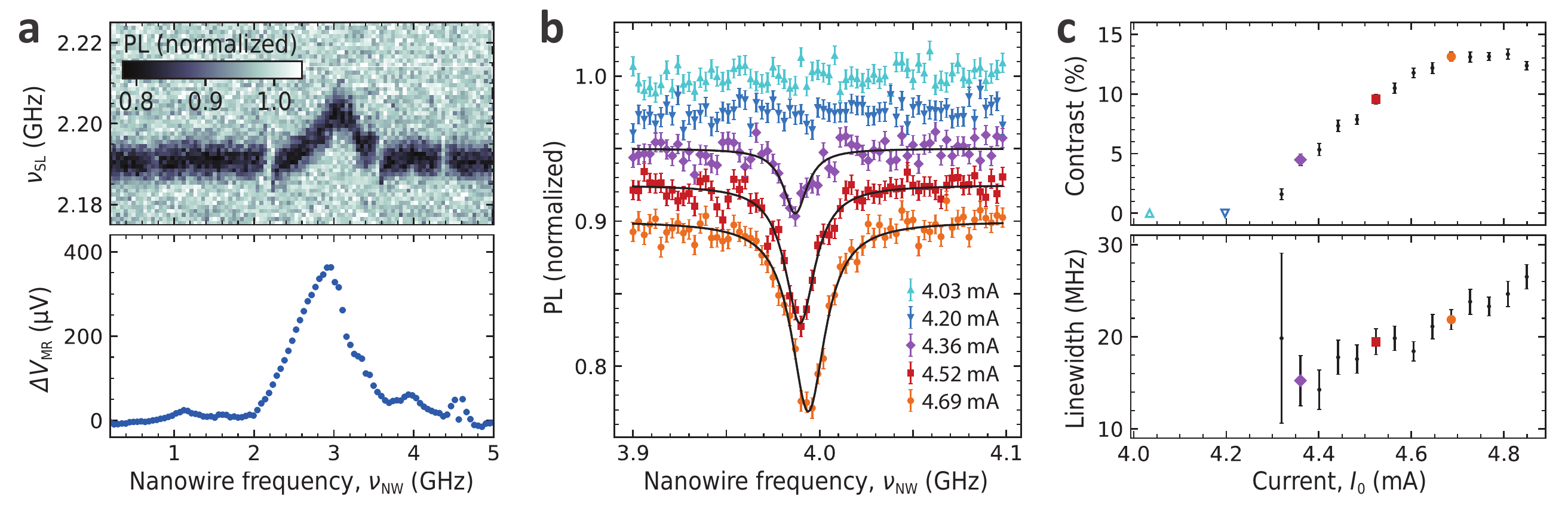}
	\caption{NV detection of parametrically-driven FMR. (a) Top: Photoluminescence (PL) versus stripline (nanowire) frequency $\nu_\text{SL}$ ($\nu_\text{NW}$) with field $\mu_0 H_0=22.5$~mT along the NV axis, and nanowire currents $I_\text{RF}=1.15\pm 0.08$~mA and $I_{0}=4.9$~mA. PL decreases near the ESR frequency $\nu_-=2.19$~GHz, and $\nu_-$ shifts due to a reduced axial stray field. Data are normalized to off-resonant levels, which is why the contrast vanishes when $\nu_\text{NW}$ matches ESR frequencies (i.e., near 2.19~GHz, 3.59~GHz) or their harmonics (4.38~GHz). Bottom: Transport FMR readout $\Delta V_\text{MR}$ under the same conditions. 
		(b) Same PL measurement as (a), but with \emph{no} stripline power, $\mu_0 H_0=29.6$~mT, and varied $I_0$, driven near $2\nu_-$. Solid curves represent Lorentzian fits used to determine the ESR contrast and linewidth.
		(c) ESR contrast (top) and linewidth (bottom) for a wider range of currents. Below $I_0=4.3$~mA, data are consistent with zero contrast (see (b)).}
	\label{fig3}
\end{figure*}

\section{Transport Characterization of Magnetic Resonances}

To verify the device's functionality, we perform ferromagnetic resonance (FMR) by applying current $I(t)=I_0 + I_\text{RF}\cos 2\pi\nu_\text{NW} t$ through the nanowire (bias $I_0$, amplitude $I_\text{RF}$, frequency $\nu_\text{NW}$), and reading the anisotropic magnetoresistance (AMR) response via a generated DC voltage $\Delta V_\text{MR}$ \cite{Liu2011Spin,supplementary}. Due to the spin Hall effect, the electrical current $I(t)$ drives a pure spin current (polarized along $\hat{y}$) into the Py layer, applying a torque $\partial_t\hat{m} \parallel \hat{y}$ on the Py magnetization's unit vector $\hat{m}=m_x\hat{x}+m_y\hat{y}+m_z\hat{z}$. The field $\vec{H}_I\parallel \hat{y}$ generated by this current applies a torque along $\hat{z}$. 

A ``typical'' ST-FMR spectrum for field $\mu_0 H_0=16.5$~mT applied along the NV axis but rotated $\theta=5^\circ$ about $\hat{z}$ is shown in Fig.~\ref{fig2}(a) (blue circles), with $I_0=0$, and $I_\text{RF}=0.95$~mA. The resonant feature at 1.6~GHz is well fit by the expected Fano-like lineshape \cite{Liu2011Spin, supplementary}, allowing us to extract the resonant frequency $\nu_\text{FMR}$ and linewidth $\Delta \nu$. Figure \ref{fig2}(b) shows the field dependence of these spectra (color scale) and the fit values of $\nu_\text{FMR}$ (points). The initial decrease in frequency corresponds to the equilibrium orientation $\langle\hat{m}\rangle$ shifting from $\hat{x}$ (where shape anisotropy provides $\nu_\text{FMR}\sim 1$~GHz) and saturating along $\approx\hat{y}$, where the anisotropy field opposes the applied field. When $\theta = 0^\circ$ (Fig.~\ref{fig2}(a), orange squares), this FMR signal vanishes (as expected above the saturation field), since both the drive torques and AMR response also vanish to lowest order in $m_x$ and $m_z$ \cite{supplementary}. Figure \ref{fig2}(c) shows the field dependence of these $\theta=0^\circ$ spectra, along with $\nu_\text{FMR}$ estimated from values at $\theta \ne 0$ \cite{supplementary}. Above saturation, $\nu_\text{FMR}$ is well fit by the Kittel formula for spatially uniform $\hat{m}$, providing effective in-plane (out-of-plane) coercive fields $\mu_0 H_{yx}=7.57\pm 0.08$~mT ($\mu_0 H_{zx} = 517\pm 4$~mT) \cite{supplementary}.

Figure~\ref{fig2}(d) shows the same $\theta = 0^\circ$ measurement with $I_0=4.08$~mA. This applies a steady torque along $-\hat{y}$ that anti-damps the magnetization ($H_I = -6 \text{ mT}$ also shifts saturation to higher fields). Importantly, a \emph{significantly} larger, asymmetric peak appears near twice the expected FMR frequency  (see Fig.~\ref{fig2}(a), red symbols), suggesting parametrically driven, large-amplitude oscillations. Simple macrospin simulations \cite{supplementary} reproduce the sign, magnitude, frequency, and line shape of this feature semi-quantitatively, and we interpret it as arising from a large-angle elliptical precession \cite{Kiselev2003Microwave} of the spatially-averaged magnetization. Within this macrospin approximation, the largest-amplitude signal (120~$\upmu$V) in Fig.~\ref{fig2}(a) corresponds to an in-plane (out-of-plane) precession angle $\approx 30^\circ$ (6$^\circ$).

%%%%%%%%%%%%%%%%%%%%%%%%%%%%%%%%%%%%%%%%%%%%%%%%
% NV Readout of Parametric Drive
%%%%%%%%%%%%%%%%%%%%%%%%%%%%%%%%%%%%%%%%%%%%%%%%

\section{NV Response to Parametric Oscillations}

This picture is validated by the NV's PL spectrum shown in Fig.~\ref{fig3}. First, while driving the nanowire at frequency $\nu_\text{NW}$, we apply a $\pi$-pulse with a nearby stripline to determine the NV’s lower electron spin resonance (ESR) frequency. Figure~\ref{fig3}(a) shows the PL response (top) and $\Delta V_\text{MR}$ (bottom) for field $\mu_0 H_0 = 22.5$~mT along the NV axis. Reductions in PL are associated with driving the NV spin from its $m_s = 0$ spin projection to $m_s = \pm 1$, occurring at resonant frequencies $\nu_\pm$. Notably, the frequency $\nu_-$ of the stripline-driven transition shifts by up to 13~MHz when $\hat{m}$ is driven to large amplitude, and follows a path qualitatively similar to $\Delta V_\text{MR}$. The maximum shift in $\nu_-$ corresponds to a decrease in the NV-axial magnetic field of 0.5~mT, as expected for the reduced average magnetization. This is a large fraction of the estimated 2.4~mT total stray field \cite{supplementary}, corresponding to an in-plane precession angle $\approx60^\circ$ that is within $10\%$ of the value (55$^\circ$) estimated from $\Delta V_\text{MR}$ in Fig.~\ref{fig3}(a).

More compellingly, we can drive the NV transition using only the Py layer's stray field. Figure \ref{fig3}(b) shows a similar set of measurements in the absence of stripline current, with $\mu_0 H_0$ tuned to 29.6~mT, such that $\nu_\text{FMR} \approx \nu_-$. Above $I_0=4.3$~mA, a PL dip (corresponding to field-driven transitions) appears at \emph{twice} $\nu_-$, as expected for parametric oscillations at half the drive frequency. Figure \ref{fig3}(c) shows the contrast and linewidth of this PL dip for a wider bias range, illustrating a sharp threshold at $I_0=4.3$~mA. This measured PL response provides unambiguous evidence of ST anti-damping and large-amplitude, parametrically driven oscillations. We note that details such as the precise threshold current and shapes of the bias dependences in (c), depend somewhat on $\vec{H}_I$, which slightly tunes the NV and FMR frequencies.

%%%%%%%%%%%%%%%%%%%%%%%%%%%%%%%%%%%%%%%%%%%%%%%%%%%
% PHASE SPACE, ODDITIES
%%%%%%%%%%%%%%%%%%%%%%%%%%%%%%%%%%%%%%%%%%%%%%%%%%%

\begin{figure}
	\centering
	\includegraphics[width=2.3in]{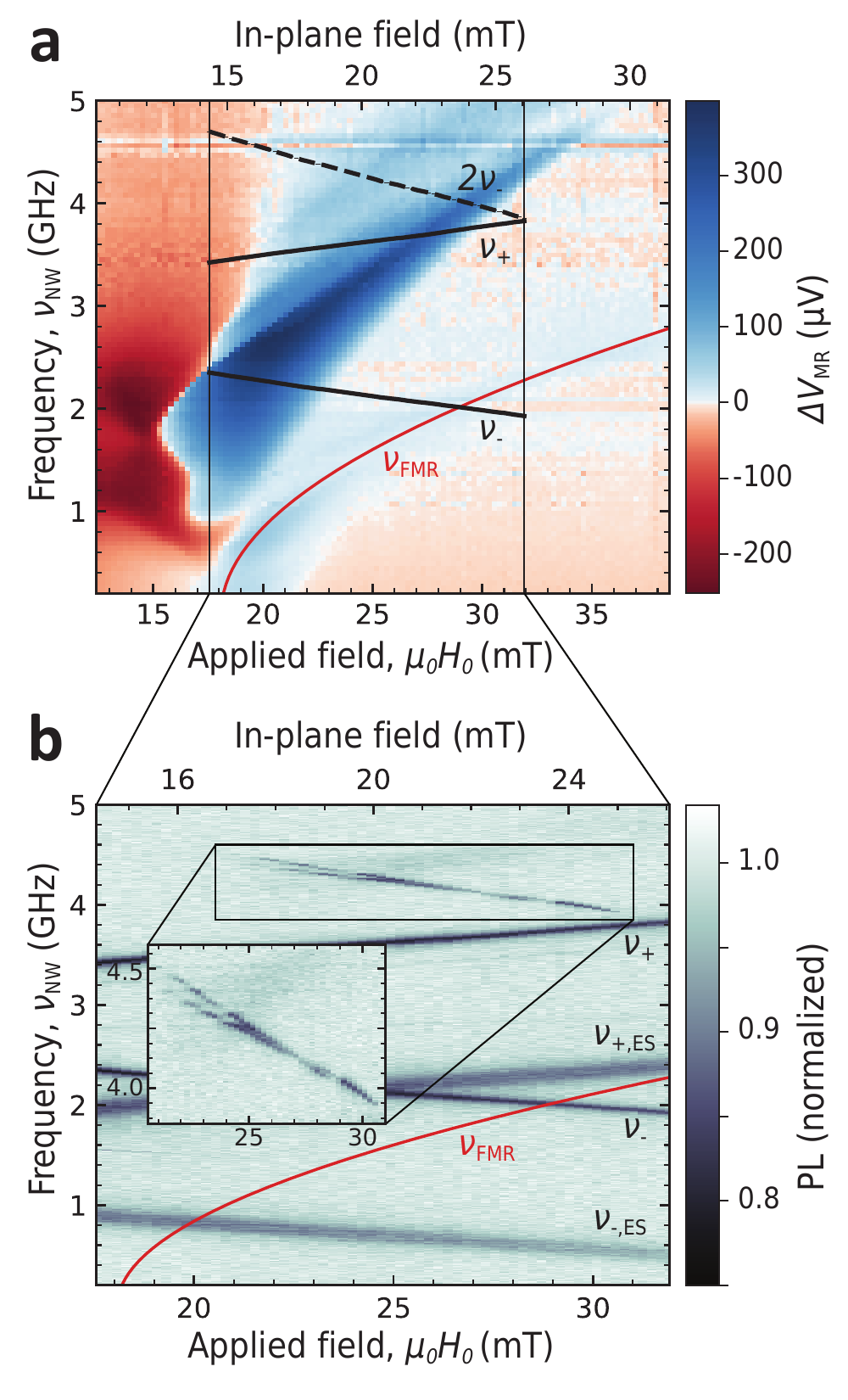}
	\caption{Phase space for parametrically driven (a) $\Delta V_\text{MR}$ and (b) PL at $I_{0}=4.9$~mA and $I_\text{RF}=1.15$~mA. The four prominent PL features correspond to current-driven ESR transitions of the ground ($\nu_\pm$) and excited ($\nu_{\pm,\text{ES}}$) states. Inset of (b) shows the Py-driven feature at $2\nu_-$. Red curves indicate the expected FMR frequency $\nu_\text{FMR}$.}
	\label{fig4}
\end{figure}

In Fig.~\ref{fig4}, we map out the parametrically driven phase space of (a) $\Delta V_\text{MR}$, and (b) PL under the same conditions as Fig.~\ref{fig3}. For this larger drive, new features appear above the primary parametric $\Delta V_\text{MR}$ peak, which we tentatively attribute to higher order spin-wave modes \cite{Duan2014Spin}. Of interest, the Py-driven PL feature near $2\nu_-$ from Fig.~\ref{fig3}(b) extends over a wide range of fields, consistent with the presence of these higher-frequency transport features. 

At this larger drive, we observe a residual directly-driven precession amplitude up to $\sim 14^\circ$ at $\nu_\text{FMR}$ in the transport data (Fig.~\ref{fig4}(a)). Despite this, we observe no evidence of spin flips at $\nu_\text{FMR}$ in the PL spectra of Fig.~\ref{fig4}(b), contrasting what is ubiquitously observed near extended films \cite{Wolfe2014Off, Van2015Nanometre, Wolfe2016Spatially, Page2016OpticallyARXIV, Du2017Control}. We tentatively attribute this to the reduced density of states in our confined geometry, which acts to suppress multi-magnon up-conversion processes.

Finally, as shown in the inset, the Py-driven spin resonance splits by up to $\sim$60~MHz at some fields. The nature of this splitting will be the subject of future work, but we suspect it is due to the presence of quasistable magnetic modes \cite{Sankey2005Mechanisms, Berkov2007Magnetization} having different average stray fields. Notably, information about these dynamics are \emph{not} apparent in (a), highlighting this technique's potential to provide qualitatively new information. 

%%%%%%%%%%%%%%%%%%%%%%%%%%%%%%%%%
% Cooling / Heating
%%%%%%%%%%%%%%%%%%%%%%%%%%%%%%%%%

\section{Strongly Damped Magnetization}

The associated NV spin relaxation rates $\Gamma_\pm$ can also probe the local field noise, providing some information about the thermal occupancy of spin wave modes~\cite{Du2017Control,Demidov2017Chemical}. Stated briefly, the noise spectral density $S_\perp$ of the stray field perpendicular to the NV axis (units of T$^2$/Hz) should have the form
\begin{equation}
S_\perp (\nu,H_0) = \sum_\mathbf{k} \bar{n}(\nu_\mathbf{k}(H_0)) P_\mathbf{k}(\nu,H_0) f_\mathbf{k},
\end{equation}
where $\bar{n}(\nu_\mathbf{k})$ is the thermal occupancy of the spin wave mode having wavenumber $\mathbf{k}$ and frequency $\nu_\mathbf{k}$, $f_\mathbf{k}$ is a constant converting magnon number to field noise power, and $P_\mathbf{k}(\nu,H_0)$ is a density function describing how this power is spread over the frequency domain (peaked at $\nu_\mathbf{k}$). In the presence of $S_\perp$, the NV spin relaxation rate $\Gamma(\nu)$ increases from its nominal value $\Gamma^0\sim 60$ Hz \cite{supplementary} to a value
\begin{equation}
\Gamma(\nu) = \Gamma^0 + \frac{\gamma_\text{NV}^2}{2}S_\perp,
\end{equation} 
where $\gamma_\text{NV}$ is the gyromagnetic ratio of the NV spin. Figure \ref{fig5}(a) shows the measured \cite{supplementary} relaxation rates $\Gamma_\pm$ for varied field $H_0$, along with $\nu_\pm$ and $\nu_\text{FMR}$ below. Qualitatively similarly to extended YIG films \cite{Du2017Control}, the largest relaxation occurs when when $\nu_\pm$ is just above $\nu_\text{FMR}$, with peak value determined by the balance of $\bar{n}$, $P_\mathbf{k}$, $f_\mathbf{k}$ (which takes the largest value when $2\pi/|\mathbf{k}|$ is comparable to the NV-wire distance), and the density of spin wave states. In contrast to Ref.~\cite{Du2017Control}, spin-transfer damping (and antidamping) in this confined metallic system should completely dominate over Joule heating (the nanowire temperature changes at most by $\sim$5 K \cite{supplementary}), enabling strong modification of the thermal fluctuations. Figure \ref{fig5}(b) shows how $\Gamma_\pm$ varies with $I_0$ (while simultaneously compensating $\vec{H}_I$ to fix $\nu_\text{FMR}$); the weaker $I_0$-generated field at the NV leads to modest shifts of $\nu_\pm$ in Fig. \ref{fig5}(d). Both relaxation rates \emph{significantly} decrease for negative current (damping), indicating a strong suppression of stray field noise, and increase for positive current (antidamping). Naively assuming these changes arise entirely from $\bar{n}(\nu_\pm)$, and that the action of damping is to modify the effective magnetic temperature $T_\text{eff}$ \cite{Demidov2017Chemical,Du2017Control}, the observed change in $\Gamma_-$ corresponds to ST cooling to $T_\text{eff}\sim 150$~K at $I_0=-4.08$~mA. However, the probe frequencies $\nu_\pm$ drift with $I_0$ and the ST damping broadens the peak in $S_\perp$. A model combining these effects with the observed field-dependence in Fig.~\ref{fig5}(a) \cite{supplementary} produces the ``spin transfer free'' relaxation rate estimates $\Gamma^\prime_\pm$ shown in Fig.~\ref{fig5}(b) and the color scale in the lower panels of (a) and (b). Importantly, the expected changes are much smaller (and have the opposite trend), suggesting that our estimate of $T_\text{eff}$ might represent an upper bound. We note that, in this geometry, we could not extract trustworthy parameters from FMR at maximum damping, highlighting the utility of the NV as an independent probe.

\section{Conclusions}

\begin{figure}
	\centering
	\includegraphics[width=3.3in]{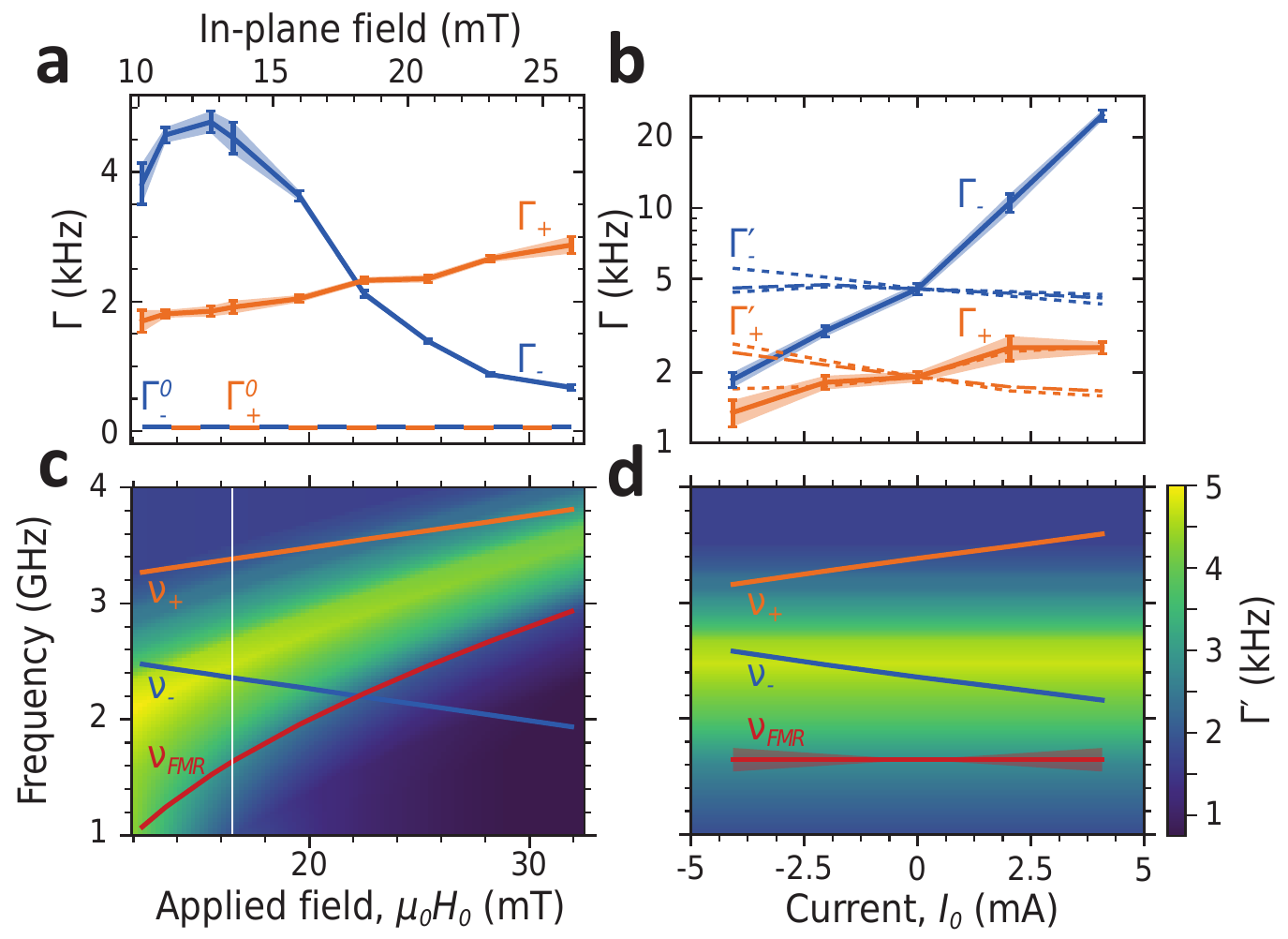}
	\caption{Spin relaxometry under bias. (a) Measured relaxation rates $\Gamma_{\pm}$ versus applied field, with bias $I_0$=0. Also shown is the internal rates $\Gamma_{\pm}^0$. (b) Measured $\Gamma_{\pm}$ for varied $I_0$ (solid lines), adjusting $\mu_0 H_0$ (nominally 16.5~mT at $I_0=0$, white line in (c)) to ensure $\nu_\text{FMR}$ remains constant to within the shown error (shading on red line in (d)). Dashed lines are estimated rates $\Gamma_\pm^\prime$ in the absence of ST effects, with absolute bounds shown as dotted lines. (c)-(d) Comparison of $\nu_{\pm}$ and $\nu_\text{FMR}$ (curves) with reconstructed estimate of $\Gamma^\prime$ (color plots) \cite{supplementary}.}
	\label{fig5}
\end{figure}

The NV PL spectrum provides qualitatively different information about magnetic nanocircuits than is provided by traditional transport measurements. We demonstrated parametric magnetic oscillations controlled by ST effects, and that ESR transitions can be driven entirely by a local magnetization. The latter technique may potentially facilitate fast, low-power control of NV spins, and the nonlinear dynamics involved could, e.g., be used to amplify incident fields~\cite{Badea2016Exploiting}. We also observe large changes in the NV spin relaxation time when subjecting the magnetic element to ST damping, consistent with magnon cooling to well below ambient temperatures. Finally, we note that the demonstrated stray field coupling between the NV and spin wave modes is quite large. Of note, Fig.~\ref{fig5}(a) exhibits a maximal $\Gamma$ at 15.5~mT, where $\nu_-$ corresponds to a room-temperature magnon occupancy $\sim$2500, meaning the noise $S_{\perp,\text{ZPF}}$ generated by the ferromagnetic zero-point fluctuations would set $\Gamma\sim$1~Hz. This value is comparable to the intrinsic relaxation rates of bulk NV's at low temperature \cite{Jarmola2012Temperature}, suggesting it might be feasible to resolve the quantum fluctuations of a magnetic nanocircuit.  

\section{Acknowledgments}
We thank Luke Hacquebard, Quentin Pognan, Isabelle Racicot, Souvik Biswas, and Tsvetelin Totev for discussions and related work. JS, LC, and M-PL acknowledge support from Fonds de Recherche - Nature et Technologies (FRQNT) PR-181274. LC acknowledges funding support from Canada Foundation for Innovation and Canada Research Chairs (CRC) project 229003, National Sciences and Engineering Research Council of Canada (NSERC) RGPIN 435554-13, and l'Institut Transdisciplinaire d'Information Quantique (INTRIQ). JS acknowledges support from CRC project 228130 \& 231509. LC and JS acknowledge Benjamin Childress and Calvin Childress for all sleep-deprived errors in judgment, including this acknowledgment.

\bibliographystyle{bibstyle-jack}
\bibliography{errthing}

\end{document}

\title{Supplementary Information}

\maketitle
\tableofcontents{}

\pagebreak

\section{Device fabrication}

We fabricate an 8-$\upmu$m-long $\times$ 417-nm-wide Py ($\mathrm{Ni_{81}Fe_{19}}$) / Pt nanowire (layer thicknesses $t_\text{Py}=10$~nm and $t_\text{Pt}=10$~nm) on CVD-grown, electronic grade diamond (Element Six), as shown in Fig.~1 of the main text. Prior to depositing the nanowire, we relieve strain and smooth the diamond surface to $\sim$0.2~nm-rms over micron length scales ($\sim$0.05~nm-rms over smaller scales) with Ar/Cl$_2$ etching \cite{Lee2008Etching}, then create a layer of NV centers 75$\pm$15~nm below the surface with $^{15}$N$^+$ implantation (60~keV, Innovion Corp.) and high-temperature annealing \cite{Chu2014Coherent}. A subsequent Ar/O$_2$ plasma etch (20 mTorr, 35 sccm Ar, 10 sccm O$_2$, 108 W RF for 45 sec; etches 20 nm/min) removes contaminants and reduces the NV depth to 60$\pm$15~nm. To remove any residual graphitized carbon and oxygen-terminate the surface, we perform a final clean in a boiling 1:1:1 mixture of nitric, perchloric, and sulfuric acid at 200$^\circ$ C for 1 hour. The Py/Pt nanowire is deposited with e-beam lithography, sputtering, and liftoff (we recommend a more directional deposition, e.g., evaporation to improve liftoff yield). We then deposit Ti (10 nm) / Au (50 nm) macroscopic leads and a microwave stripline (10~$\upmu$m wide, 17~$\upmu$m center-to-center distance) using photolithography, and finally connect the leads to the nanowire with another e-beam liftoff (Ti (10~nm) / Au (70~nm) evaporated). The contacts leave an ``active'' region of length $5.2$~$\upmu$m exposed.

\section{NV measurements}
\subsection{Single NV identification and isolation}
\label{g2}

We focus on an NV center (``NV$_A$'') having the strongest stray field coupling. However, this NV is positioned near a second NV ``NV$_B$'', that provides an additional small photoluminescence (PL) signal. As described below, we identify these two NVs using a combination of electron spin resonance (ESR) spectroscopy and photon correlation measurements, taking advantage of their different symmetry axes to isolate signals from NV$_A$. 

\begin{figure}[h!]
	\center
	\includegraphics[width=5 in]{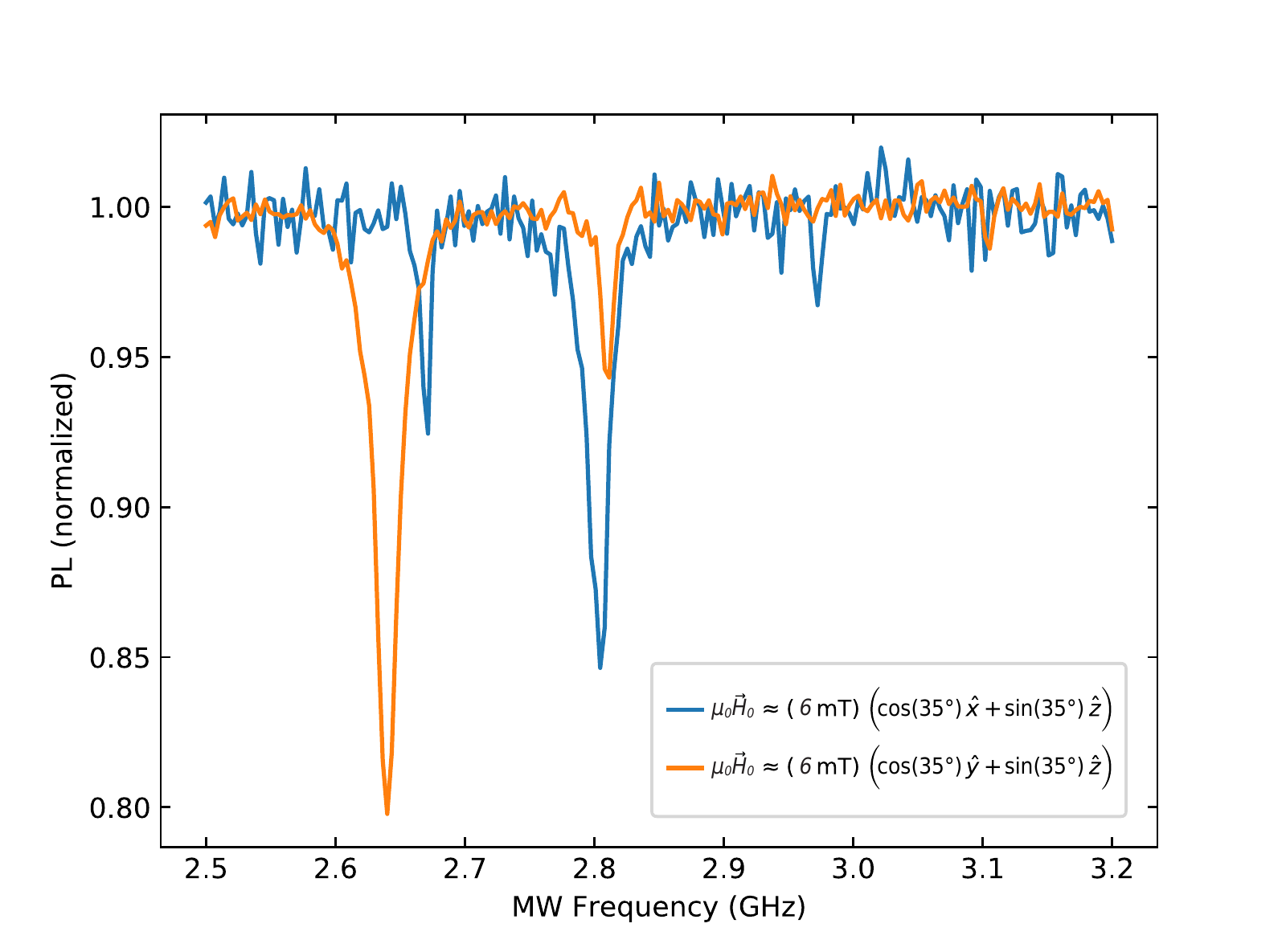}
	\caption{ESR measurement of the probed spot in Fig.~1(b) of the main text, showing PL dips associated with two NV centers.}
	\label{Fig:ESR_AB}
\end{figure}

The two NVs are both associated with the single bright emission spot shown in Fig.~1(b) of the main text. With the 532 nm excitation focused on this spot, we observe ESR spectra indicating the presence of NVs having two different orientations.  Figure ~\ref{Fig:ESR_AB} shows two ESR spectra taken with a 6~mT magnetic field oriented 35$^\circ$ from the diamond surface in the $xz$ and $yz$ planes, such that it is aligned with one of the allowed $<$111$>$ crystallographic orientations of the diamond bonds (i.e., along one of the allowed NV symmetry axes). Dips in PL are associated with driving the NV from its $m_s=0$ spin projection to $m_s=\pm 1$, occurring at frequencies $\nu_\pm$ (nominally $\nu_+=\nu_-=2.87$~GHz at zero field). The deepest photoluminescence dip in both spectra is associated with $m_s=0 \rightarrow -1$ transition of the better-coupled NV center (NV$_A$), mostly because the polarization of the 532 nm excitation is perpendicular to the NV$_A$ symmetry axis. This transition shows the largest magnetic field dependence when the field is in the $yz$ plane, parallel to NV$_A$ (corresponding to the arrow in Fig.~1(a) of the main text), while the shallower $m_s = 0 \rightarrow -1$ transition of NV$_B$ shows the largest deviation with the field aligned in the $xz$ plane. Note that the $m_s = 0 \rightarrow +1$ transition of NV$_A$ is barely visible above the noise in each spectrum, due to imperfect power coupling to our stripline at the higher frequencies (this is also why we do not see the $m_s = 0 \rightarrow +1$ transition for NV$_B$ in these spectra). In principle, there could be multiple NVs creating the observed spectra, but the total photon count rate is approximately what we expect for two NVs. Moreover, because the emission spot is close to the magnetic nanowire, in a region of high magnetic field gradient, NVs with the same orientation would be expected to show distinct ESR lines. We see no evidence of unexpected dips under any conditions, lending additional credence to the identification of two NVs. 

To further verify our interpretation that only two NV centers are co-located within the emission spot, we perform photon correlation measurements to estimate the number of contributing emitters. The second order correlation function $g^{(2)}(\tau)$ was measured with a Hanbury-Brown-Twiss setup, using a fiber beam-splitter and two single photon counting modules. A time-correlated single photon counting system (PicoHarp 300) was used to measure the histogram of times between two consecutive photons. Such a normalized distribution can be interpreted as a measurement of $g^{(2)}(\tau)$, as long as $\tau$ is much smaller than the mean time between two detection events (for us, $\tau \ll$ 1/count rate= $(30$~kcounts/s$)^{-1}$ $\approx$ 33 $\upmu$s) \cite{Kurtsiefer2000Stable}. We measure a background count rate (on a dark area of the sample) of $\approx$ 3~kcounts/s, which we subtract from our data before analysis.

\begin{figure}[h!]
	\center
	\includegraphics[width=3.5 in]{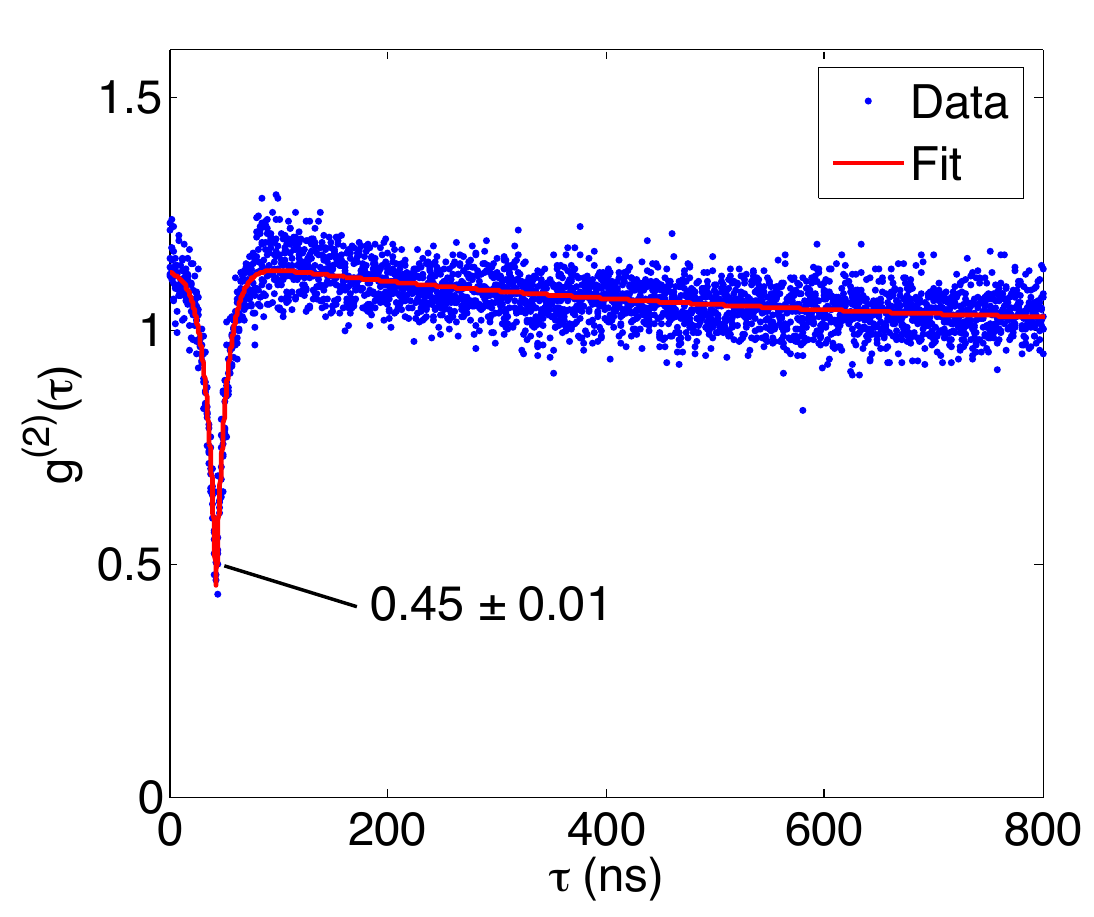}
	\caption{$g^{(2)}(\tau)$ correlation measurement of the studied emission spot, with $g^{(2)}(0)$=0.45$\pm$0.01.}
	\label{Fig:g2}
\end{figure}

Our data shows a clear anti-bunching signature at short times (see Fig.~\ref{Fig:g2}), characteristic of a small number of quantum emitters, and slight bunching at longer times, associated with metastable states in the emitters. To quantify $g^{(2)}(0)$ requires that we appropriately normalize the data. Merely setting the long-time value to 1 is problematic because of the exponential decay associated with the fact that we are not directly measuring the correlation function but a histogram of times between photons. Indeed, we observed such an exponential decays for $g^{(2)}(\tau)$ measured with attenuated laser light. We thus compensate for the decay by fitting an exponential with a decay constant $\tau$, $e^{-t/\tau}$ to the data from 1 $\mu$s onward. Multiplying the full spectrum of data points by $e^{t/\tau}$ corrects for the exponential decay, and allows us to used the long-time value for our normalization. %In order to correct for the BG we subtract the BG traces from the data. \\
We then use a three level model to fit the curve to the function \cite{Verberk2003Photon}:
\begin{equation}
g^{(2)}(\tau)=c_1\left(1+c_2e^{-\frac{\tau}{\tau_1}}+c_3e^{-\frac{\tau}{\tau_2}}\right),
\end{equation}
where c$_{1,2,3}$ are scaling factors and $\tau_{1,2}$ correspond to the lifetimes of the involved NV states. We find that $g^{(2)}(0)=1 + c_2+c_3 = 0.45\pm$0.01, consistent with two co-located emitters, where one is slightly better coupled to our optical excitation/collection path than the other.

In our measurements, we isolate NV$_A$ from NV$_B$ by taking advantage of their different orientations. For ESR measurements (such as in Fig.~3 of the main text), the transitions of NV$_A$ and NV$_B$ are spectrally resolved. Moreover, we align the polarization of the 532 nm illumination to preferentially excite NV$_A$, reducing photoluminescence from NV$_B$; we also work in magnetic fields aligned with NV$_A$, which reduces the ESR contrast of NV$_B$. As a result, the resonances from NV$_B$ are only weakly visible in the spin resonance data shown in Fig.~4(b) of the main text. For spin relaxation measurements, we use $\pi$ pulses tuned to the transitions of NV$_A$ to isolate its relaxation rates from those of NV$_B$, as described in further detail below. 

\subsection{Extracting spin relaxation rates}

We extract the two relaxation rates $\Gamma_{\pm}$ for NV$_A$ by initializing the spin into the three spin projections $m_s = -1,0,1$, observing its relaxation via photoluminescence measurements, and fitting our data to a rate equation model. 

We prepare the spin projections using a pulse of 532 nm excitation,  which optically pumps both NV$_A$ and NV$_B$ into $m_s = 0$; to prepare $m_s = \pm 1$, we then apply a $\pi$ pulse of microwaves tuned to the NV$_A$ spin transitions for durations of 50-200 ns, chosen to maximize the contrast of a Rabi sequence. These $\pi$ pulses are far off resonance with NV$_B$, so it remains in $m_s = 0$. After a variable delay of duration $\tau$, we again apply a pulse of 532 nm excitation and observe photoluminescence counts, which are higher for $m_s = 0$ states than for $m_s = \pm 1$. In the end, we record three photoluminescence traces: $f_0(\tau), f_+(\tau)$, and $f_-(\tau)$, where the subscript indicates the spin state into which NV$_A$ was initialized. 

Since these photoluminescence traces include emission from both NV$_A$ and NV$_B$, we consider the differences in the photoluminescence traces $d_-(\tau) = f_0(\tau) - f_-(\tau)$ and $d_+(\tau) = f_0(\tau) - f_+(\tau)$. Because NV$_B$ is prepared in the same way for all three initializations, its (possibly $\tau$-dependent) photoluminescence is eliminated in the data sets $d_\pm(\tau)$, as is any background photoluminescence not associated with NV$_A$.  

We then simultaneously fit $d_-(\tau)$ and $d_+(\tau)$ to a rate equation model for spin relaxation:
\begin{eqnarray}
\frac{d p_-(t)}{dt} &=& \Gamma_- \left(p_0(t) -p_-(t)\right)\nonumber\\ 
\frac{d p_+(t)}{dt} &=& \Gamma_+\left(p_0(t) -p_+(t)\right)\nonumber\\
\frac{d p_0(t)}{dt} &=& -(\Gamma_- + \Gamma_+) p_0(t)  + \Gamma_- p_-(t) + \Gamma_+ p_+(t),
\end{eqnarray}
where $p_m(t)$ is the population of NV$_A$ in spin state $m$ and $\Gamma_\pm$ are the relaxation rates on the $m_s = 0 \leftrightarrow m_s = \pm 1$ transitions. We also considered double quantum relaxation between $m_s = \pm 1$ states, but our fits revealed that this rate was insignificant. The fit functions for $d_\pm(\tau)$ are found by calculating the population in $m_s = 0$ for three different spin initializations,  $p_0(\tau)^{(m)}$ (where the superscript indicates the initial state), and subtracting to reconstruct a signal proportional to $d_\pm(\tau) \propto p_0(\tau)^{(0)} - p_0(\tau)^{(\pm)}$. Note that this proportionality holds even for imperfect optical pumping and imperfect $\pi$ pulses -- low fidelity initialization and incomplete $\pi$ pulses reduce the contrast of our signals, but not their time dependence. We can thus fit $d_\pm(\tau)$ to $A \left( p_0(\tau)^{(0)} - p_0(\tau)^{(\pm)}\right)$, where $A$ is an unknown fitting parameter, and extract the two relaxation rates $\Gamma_{\pm}$.

\section{Magnetic field calibration}

We apply static magnetic fields using a permanent magnet (K\&J magnetics, D42-N52) mounted on a three-axis translation stage equipped with closed-loop motorized actuators (CONEX-TRB25CC). The 25mm range of motion permits us to apply magnetic fields up to 90 mT oriented along the NV symmetry axis. 

To calibrate the relationship between the actuator position and the applied magnetic field, we exploit the field-dependent photoluminescence of the NV center~\cite{Tetienne2012Magnetic}. Essentially, NV photoluminescence is maximized when the magnetic field is parallel to its symmetry axis, with heightened sensitivity near the ground- and excited-state level anti-crossings near 103 mT and 51 mT, respectively. Following the procedure described in Ref.~\cite{Van2015Nanometre}, we measure the photoluminescence of an NV as a function of the position of the permanent magnet. From this data, we can extract the controller coordinates that bring the NV center to the excited state level anti-crossing (51 mT) and align the magnetic field with the NV axis. We then model the stray field of the permanent magnet using RADIA \cite{Elleaume1997Computing} and determine the position (relative to the permanent magnet origin) where the stray field is 51 mT aligned with the NV axis. This provides a coordinate transformation between our controller position and the model. We can then use the RADIA model to roughly predict the field as a function of controller position. We design a trajectory for the magnet that should maintain the magnetic field along the NV axis or at a defined angle relative to that axis. 

\begin{figure}[h!]
	\center
	\includegraphics[width=6in]{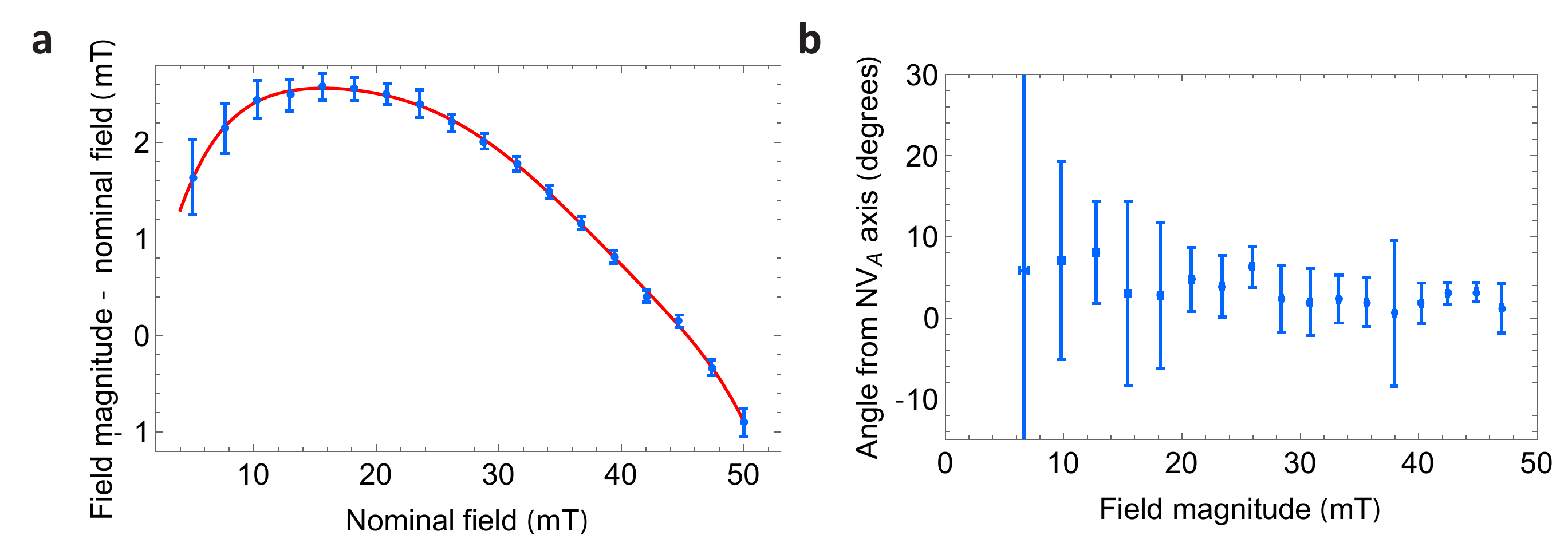}
	\caption{(a) Magnetic field magnitude (as determined by ESR measurements), relative to the expected magnetic field (as calculated by RADIA). Red curve is a sixth order polynomial fit used to rescale the magnetic field values in the main text. (b) Off-axis angle of the magnetic field as a function of magnitude. The (large) error bars represent systematic uncertainties in our system parameters.}
	\label{Fig:magnetCalibration}
\end{figure}

This calibration procedure gives only an approximation of the magnetic field due to potential inaccuracy of the RADIA model and possible misalignment between the (nominally aligned) coordinate axes of our sample and the controller axes. We therefore use NV electron spin resonance (ESR) measurements to determine the actual magnetic field taken along the trajectories used in our experiments. Note that these measurements were taken after the device magnetization disappeared (see Sec.~\ref{Sec:Bstray}), so the NV experiences only the applied magnetic field. We fit the ESR spectra to extract the transition frequencies $\nu_\pm$. From the linear and quadratic Zeeman effects, we can determine the on- and off-axis components of the magnetic field \cite{Van2015Nanometre}, and thus calculate its magnitude. Figure~\ref{Fig:magnetCalibration}(a) shows the difference between the extracted and expected field magnitudes as a function of the expected field predicted by the RADIA model. These values are found using an NV g-factor of 2.0030(3) and a zero field splitting of 2.870(2) GHz, consistent with our observed splittings and literature values \cite{Doherty2013The, Felton2009Hyperfine}. The solid red curve is a sixth order polynomial interpolating function used to rescale the magnitudes of the magnetic fields in the data presented in the main text. Figure~\ref{Fig:magnetCalibration}(b) shows the extracted angle of the magnetic field, which is poorly constrained by these measurements; due to the good ESR contrast observed over the entire magnetic field scan at $\theta = 0^\circ$ in Fig.~4(b) of the main text, and the strong suppression of directly-driven FMR at $\theta = 0^\circ$, we believe the angular alignment is good to within a couple of degrees. 

% For all $\theta = 0^\circ$ data shown in the main text, we give the magnetic field using the {\color{red} component or magnitude???} of the magnetic field along NV$_A$ as measured via ESR of NV$_A$ after the magnetization of our device was determined to have vanished (possibly due to oxidation or electromigration effects from strong applied currents). We verified that these ESR spectra were indistinguishable from those taken on an NV of the same orientation far from the device.  Along the trajectory taken, the angle of the magnetic field varies slightly; for magnetic fields nominally applied along the NV$_A$ axis, the angle is maintained within ???$^\circ$ of the desired orientation (see Fig.~\ref{angle}). Based on observed drifts in sample position over time, we estimate that the magnetic field is accurate to within approximately ??? \%. 

%{\color{red} FIGURE NEEDED: ANGULAR VARIATION OF THE MAGNETIC FIELD ALONG THE TRAJECTORY || NVA \label{angle}}

\pagebreak
\section{Spin transfer control and magnetoresistive readout}\label{sec:transport}

We control the magnetization of the Py layer with current $I(t)$ in the nanowire, which provides two torques. First, as indicated in Fig.~1(a) of the main text, the current density $J_\text{Pt}$ in the Pt layer produces a spin current density $J_\text{s} = \theta_\text{SH} J_\text{Pt}$ (where $\theta_\text{SH}$ is Pt's spin Hall angle), polarized along $\hat{y}$ and travelling along $\hat{z}$ \cite{Valenzuela2006Direct}. A fraction $\eta$ of these spins are absorbed by the Py layer ($\eta\theta_\text{SH}=0.055$ \cite{Liu2011Spin}), thereby applying an areal spin transfer torque (STT) density at the Py interface, or an effective per-volume torque density (N$\cdot$m/m$^3$)
\begin{equation}
\vec{\mathscr{T}}_{\text{STT}}=\frac{\hbar \eta\theta_{\text{SH}}J_{\text{Pt}}}{2 e t_\text{Py}}\hat{m}\times\left(\hat{m}\times\hat{y}\right),
\end{equation}
where the unit vector $\hat{m} = m_x \hat{x}+m_y\hat{y}+m_z\hat{z}$ represents the orientation of the local magnetization. Second, the current in the wire generates a magnetic field $\vec{H}_I$ in the Py layer, with a torque (volume) density
\begin{equation}
\vec{\mathscr{T}}_{H_I}=\mu_{0}M_{s}\vec{H}_I\times\hat{m},
\end{equation}
where $M_s$ is the Py saturation magnetization. Changes in the magnetization can then be read out via the device's resistance, which varies approximately\footnote{This assumes $m_z\ll m_x$, which is justified by the calculations and simulations in Sec.~\ref{sec:macrospin}.} as
\begin{equation}\label{eq:R}
R= R_0 (1+\delta_\text{AMR}m_x^2),
\end{equation}
where $R_0 = 237~\Omega$ is the wire's unbiased resistance at $m_x=0$, and $R_0\delta_\text{AMR}\approx0.2~$m$\Omega$ is the change due to anisotropic magnetoresistance (AMR) from the fraction of current flowing through the Py layer (measured by monitoring $R$ while sweeping the field). 

To verify device functionality, we perform spin transfer ferromagnetic resonance (ST-FMR) similar to that of Ref.~\cite{Liu2011Spin}. To briefly summarize, a current $I(t) = I_0 + I_\text{RF}\cos(2\pi\nu_\text{NW} t)$ (with DC bias $I_0$, ``radio frequency'' (RF) amplitude $I_\text{RF}$, and frequency $\nu_\text{NW}$) driving coherent resistance oscillations $R(t) = R_0 + \Delta R_\text{0} + \Delta R_\text{RF}\cos(\nu_\text{NW} t+\psi)$ (for constant offset $\Delta R_0$, amplitude $\Delta R_\text{RF}$, and phase $\psi$) will generate a time-averaged voltage change
\begin{equation}\label{eq:DeltaV}
\Delta V = I_0 \Delta R_0 + \frac{1}{2}I_\text{RF} \Delta R_\text{RF}\cos{\psi}
\end{equation}
that can be read out with a lock-in technique (see Sec.~\ref{subsec:Calibration-technique}). This voltage comprises a resonant magnetoresistance signal $\Delta V_\text{MR}$ (estimated for a macrospin in Sec.~\ref{sec:macrospin}) and a broad background due to Joule heating (useful for estimating $I_\text{RF}$, as discussed in Sec.~\ref{sec:joule-background}) that is subtracted from all presented spectra.

\section{Joule heating background, differential resistance, and RF current calibration}\label{sec:joule-background}

When a time-dependent current passes through a resistive wire, it dissipates a time-dependent power, thereby heating the sample and causing a time-dependent change in resistance. In general, this nonlinear response can generate a static voltage $\langle V\rangle$ across the resistor. Here we develop a framework for quantifying this effect, and show how it can be used to provide a reasonable estimate of the RF current flowing through our nanowires. This calibration considers RF temperature oscillations not included in (e.g.) Tshitoyan \emph{et al} \cite{Tshitoyan2015Electrical}, which are especially relevant when the substrate is highly thermally conductive.

\subsection{Derivation of Joule heating mixdown voltage}\label{subsec:Derivation-of-Joule}

In general, the instantaneous voltage $V$ will depend on current
$I$ and the change in temperature $\Delta T$. For small $I$ and
$\Delta T$, we can Taylor expand to 3rd order:
\begin{align}
V & \approx\left(\partial_{I}V\right)I+{\color{teal}\left(\partial_{T}V\right)\Delta T}\\
& +{\color{red}\frac{1}{2}\left(\partial_{I}^{2}V\right)I^{2}}+\left(\partial_{T}\partial_{I}V\right)I\Delta T+{\color{teal}\frac{1}{2}\left(\partial_{T}^{2}V\right)\Delta T^{2}}\\
& +{\color{red}\frac{1}{6}\left(\partial_{I}^{3}V\right)I^{3}}+{\color{red}\frac{1}{2}\left(\partial_{T}\partial_{I}^{2}V\right)I^{2}\Delta T}+{\color{blue}\frac{1}{2}\left(\partial_{T}^{2}\partial_{I}V\right)I\Delta T^{2}}+{\color{teal}\frac{1}{6}\left(\partial_{T}^{3}V\right)\Delta T^{3}},
\end{align}
where all of the parenthetical factors are constants of the system
determined by the resistor's geometry, materials, and thermal anchoring.
We simplify this with the following assumptions:
\begin{enumerate}
	\item In the absence of temperature change, the resistor responds linearly.
	This means the terms \textcolor{red}{(red)} having no temperature dependence $\partial_{I}^{2}V=\partial_{I}^{3}V=0$; as a result, $\partial_{T}\partial_{I}^{2}V=0$ as well. 
	\item Changing the temperature does not on its own generate a voltage. This
	means the current-independent terms $\partial_{T}V=\partial_{T}^{2}V=\partial_{T}^{3}V=0$
	\textcolor{teal}{(teal)}.
	\item The temperature change $\Delta T\sim I^{2}$ to lowest order, meaning
	the penultimate term \textcolor{blue}{(blue)} is of order $I^{5}$ and can be dropped.\footnote{In systems having a more significant Peltier effect (e.g., some spin
		valves), this should not be dropped.}
\end{enumerate}
Under these assumptions,
\begin{equation}
V \rightarrow \left(\partial_{I}V\right)I+\left(\partial_{T}\partial_{I}V\right)I\Delta T.\label{eq:instantaneous-voltage}
\end{equation}

The first term is the heat-free linear response of the resistor, with
the constant $\partial_{I}V$ being the resistance near $I=0$. We
emphasize that $\partial_{I}V$ is a constant that does \emph{not
}depend on $I$ or $\Delta T$ (all differentials in Eq.~\ref{eq:instantaneous-voltage}
are evaluated at $I=\Delta T=0$) and is qualitatively different
from laboratory measurements commonly referred to as ``differential resistance'', wherein the current is slowly modulated and the resulting
voltage modulations are recorded (see Section \ref{subsec:Calibration-technique}). 

The second term is the Joule heating nonlinearity of interest; an
``extra'' voltage can be generated from this term only if there
is both a current \emph{and} a temperature change. For example, if
$I=I_{0}$ is some constant value, this will heat the sample, causing
$\Delta T>0$, which raises the resistance by $\left(\partial_{T}\partial_{I}V\right)\Delta T$,
at which point $I_{0}$ produces an additional voltage. If $I$ changes
slowly enough with time that the system remains in steady state, it
is this term that is responsible for a bias dependence in a low-frequency
differential resistance measurement (see Sec.~\ref{subsec:Traditional-lock-in-readout}).

We eliminate temperature $\Delta T$ by assuming small enough changes
that $\Delta T$ responds linearly to the applied power $P$. In this
limit, an oscillatory component in the power $P_{1}\cos\omega t$
of amplitude $P_{1}$ and frequency $\omega$ will induce a temperature
change
\begin{equation}
\Delta T=\left[X(\omega)\cos\omega t+Y(\omega)\sin\omega t\right]P_{1},
\end{equation}
where $X(\omega)$ and $Y(\omega)$ (units of K/W) represent the thermal
transfer function of the wire, capturing the magnitude and phase shift
of the thermal response.\footnote{We choose this quadrature formulation of the transfer function over
	the ``usual'' complex one due to the nonlinear operations in the
	rest of the analysis. The quadrature basis is also very convenient
	for calculating the mixdown voltage, which requires only the in-phase
	component $X(\omega)$.} In our experiments, we apply a current of the form
\begin{equation}
I=I_{0}+I_{1}\cos\omega t,
\end{equation}
for constants $I_{0}$, $I_{1}$ and frequency $\omega$, such that
the instantaneous power is, to leading order (again assuming negligible
Peltier effects)
\begin{align}
P(t) & \approx\left(I_{0}+I_{1}\cos\omega t\right)^{2}\left(\partial_{I}V\right)\nonumber \\
& =\left(I_{0}^{2}+\frac{1}{2}I_{1}^{2}+2I_{0}I_{1}\cos\omega t+\frac{1}{2}I_{1}^{2}\cos2\omega t\right)\left(\partial_{I}V\right).\label{eq:Power-oscillations}
\end{align}
This comprises a thermal drive at three frequencies (zero, $\omega$,
and $2\omega$) which, in this linear-response limit, can be treated
separately. As such, the temperature
\begin{align}
\Delta T & \approx P_{0}X_{0}+P_{1}X(\omega)\cos\omega t+P_{1}Y(\omega)\sin\omega t+P_{2}X(2\omega)\cos2\omega t+P_{2}Y(2\omega)\sin2\omega t\label{eq:temperature-change}
\end{align}
with
\begin{align}
P_{0} & \equiv\left(I_{0}^{2}+\frac{1}{2}I_{1}^{2}\right)\left(\partial_{I}V\right)\\
P_{1} & \equiv\left(2I_{0}I_{1}\right)\left(\partial_{I}V\right)\\
P_{2} & \equiv\left(\frac{1}{2}I_{1}^{2}\right)\left(\partial_{I}V\right).
\end{align}
Plugging Eq.~\ref{eq:temperature-change} into the voltage expansion
of Eq.~\ref{eq:instantaneous-voltage}, 
\begin{align}
V & \approx\left(\partial_{I}V\right)\left(I_{0}+I_{1}\cos\omega t\right)\\
& +\left(\partial_{T}\partial_{I}V\right)\left(I_{0}+I_{1}\cos\omega t\right)\left(P_{0}X(0)\right)\\
& +\left(\partial_{T}\partial_{I}V\right)\left(I_{0}+I_{1}\cos\omega t\right)\left(P_{1}X(\omega)\cos\omega t+P_{1}Y(\omega)\sin\omega t\right)\\
& +\left(\partial_{T}\partial_{I}V\right)\left(I_{0}+I_{1}\cos\omega t\right)\left(P_{2}X(2\omega)\cos2\omega t+P_{2}Y(2\omega)\sin2\omega t\right).
\end{align}
Taking a time-average yields 
\begin{align}
\langle V\rangle & =\left(\partial_{I}V\right)I_{0}+\left(\partial_{T}\partial_{I}V\right)\left(I_{0}P_{0}X(0)+\frac{1}{2}I_{1}P_{1}X(\omega)\right)\nonumber \\
& =I_{0}\left(\partial_{I}V\right)\left(1+\chi(0)I_{0}^{2}+\frac{1}{2}\left[\chi(0)+2\chi(\omega)\right]I_{1}^{2}\right)\label{eq:V}
\end{align}
where we have defined a fractional resistance change transfer function
\begin{equation}
\chi(\omega)\equiv\left(\partial_{T}\partial_{I}V\right)X(\omega)
\end{equation}
for brevity (units of 1/mA$^{2}$). The first term is Ohm's law and
the second is the steady state heating due to DC current. The third
term is the response to an RF drive, and is responsible for large
backgrounds observed in FMR (see Sec.~\ref{subsec:Single-shot-lock-in-measurement}).
Importantly, this comprises two terms, the first $(\chi(0))$ arising
from the time-averaged power absorbed by the wire, and the second
($\chi(\omega))$ arising from the heat-induced resistance oscillations
at $\omega$ mixing with the drive current. 

\subsection{Differential resistance and estimation of RF current} \label{subsec:Calibration-technique}

We now discuss a means of interpreting the signal from a lock-in-based
differential resistance measurement in the presence of DC and RF current.
Section \ref{subsec:Traditional-lock-in-readout} discusses a ``traditional''
continuous-wave lock-in approach, and Sec.~\ref{subsec:Single-shot-lock-in-measurement}
discusses an impulse method that happens to be easier with our apparatus.
Section \ref{subsec:Estimating-RF-current} then discusses a means
of using the observed voltages in these measurements to estimate the
RF current flowing through the device.

\subsubsection{Traditional lock-in readout}\label{subsec:Traditional-lock-in-readout}

When ``differential resistance'' is measured using a lock-in technique,
the bias is modulated, such that the current 
\begin{equation}
I_{0}\rightarrow I_{0}+I_{m}\cos\omega_{m}t,
\end{equation}
where $I_{m}$ is the lock-in's modulation amplitude, and $\omega_{m}\ll\omega$
is the modulation frequency (typically chosen to be low enough that
the device remains in steady state). In this case, the voltage (Eq.~\ref{eq:V})
becomes
\begin{align}
\langle V\rangle & \rightarrow\left(\partial_{I}V\right)\left(I_{0}+I_{m}\cos\omega_{m}t\right)\left(1+\chi(0)\left(I_{0}+I_{m}\cos\omega_{m}t\right)^{2}+\frac{1}{2}\left[\chi(0)+2\chi(\omega)\right]I_{1}^{2}\right)\\
& =...+\left(\partial_{I}V\right)\left(I_{LI}+\chi(0)\left[3I_{0}^{2}I_{m}+\frac{3}{4}I_{LI}^{3}\right]+I_{m}\left[\frac{1}{2}\chi(0)+\chi(\omega)\right]I_{1}^{2}\right)\cos\omega_{m}t
\end{align}
where we have lumped all terms not contributing to the measured amplitude
at $\omega_{m}$ into ``...'' for brevity, and used the identity
\begin{equation}
\cos^{3}\omega_{m}t=\frac{3}{4}\cos\omega_{m}t+\frac{1}{4}\cos3\omega_{m}t.
\end{equation}
The lock-in measurement demodulates at $\omega_{m}$ to record the
response amplitude, which is then divided by $I_{m}$ to define a
``differential resistance''
\begin{equation}
R_{\text{diff}}\equiv\left(\partial_{I}V\right)\left(1+\chi(0)\left[3I_{0}^{2}+\frac{3}{4}I_{m}^{2}\right]+\left[\frac{1}{2}\chi(0)+\chi(\omega)\right]I_{1}^{2}\right).\label{eq:Rdiff}
\end{equation}
Note this result agrees with Eq.~\ref{eq:V} in the low-frequency
limit $\chi(\omega)\rightarrow\chi(0)$, but here we can see how the
presence of RF current will alter this signal.

\subsubsection{Impulse readout}\label{subsec:Single-shot-lock-in-measurement}

In our experiment, we employ an impulse readout of $R_{\text{diff}}$
wherein the amplitude of the modulation is slowly increased to its
maximum value and then decreased to zero as
\begin{align}
I_{0} & \rightarrow I_{0}+I_{s}(t)\\
I_{s}(t) & \equiv I_{m}\sin^{2}\left(\frac{\omega_{m}}{2N}t\right)\sin\left(\omega_{m}t\right)\\
& =\frac{I_{m}}{4}\left(2\sin\left(\omega_{m}t\right)-\sin\left[\omega_{m}\left(1+\frac{1}{N}\right)t\right]+\sin\left[\omega_{m}\left(1+\frac{1}{N}\right)t\right]\right),
\end{align}
where $N$ is a large integer. We have expanded $I_{s}$ in the last
line to highlight that this waveform includes only 3 frequencies near
$\omega_{m}$, which helps minimize artifacts associated with abrupt
changes in current. Similar to the continuous-wave measurement, the
device responds adiabatically for sufficiently small $\omega_{m}$,
and the low-frequency voltage becomes
\begin{align}
\langle V\rangle&\rightarrow V_{0}+V_{s}(t)\\
 & =\left(\partial_{I}V\right)\left(I_{0}+I_{s}(t)\right)\left(1+\chi(0)\left(I_{0}+I_{s}(t)\right)^{2}+\left[\frac{1}{2}\chi(0)+\chi(\omega)\right]I_{1}^{2}\right)\nonumber \\
& =\left(\partial_{I}V\right)\left(\text{DC}+\left(1+3\chi(0)I_{0}^{2}+\left[\frac{1}{2}\chi(0)+\chi(\omega)\right]I_{1}^{2}\right)I_{s}+3\chi(0)I_{0}I_{s}^{2}+\chi(0)I_{s}^{3}\right),\label{eq:V0+Vs}
\end{align}
where ``DC'' contains all time-independent terms. We then extract
a similar ``differential resistance'' $R_{\text{diff,s}}$ by taking
an overlap with the injected impulse $I_{s}(t)$ to extract a (normalized)
``in-phase'' response amplitude
\begin{align}
R_{\text{diff,s}} & =\frac{1}{I_{m}}\times\frac{\int_{0}^{2\pi N/\omega}\left[V_{0}+V_{s}(t)\right]I_{s}(t)dt}{\int_{0}^{2\pi N/\omega}I_{s}^{2}(t)dt}\\
& =\frac{8}{3\pi NI_{m}^{3}}\int_{0}^{2\pi N/\omega}V_{s}(t)I_{s}(t)dt.
\end{align}
By symmetry, the even powers of $I_{s}$ in Eq.~\ref{eq:V0+Vs} vanish,
leaving behind
\begin{equation}
R_{\text{diff,s}}=\left(\partial_{I}V\right)\left(1+\chi(0)\left(3I_{0}^{2}+\frac{35}{64}I_{m}^{2}\right)+\left[\frac{1}{2}\chi(0)+\chi(\omega)\right]I_{1}^{2}\right).\label{eq:Rdiffs}
\end{equation}

\begin{figure}[h!]
	\noindent \begin{centering}
		\includegraphics[width=9cm]{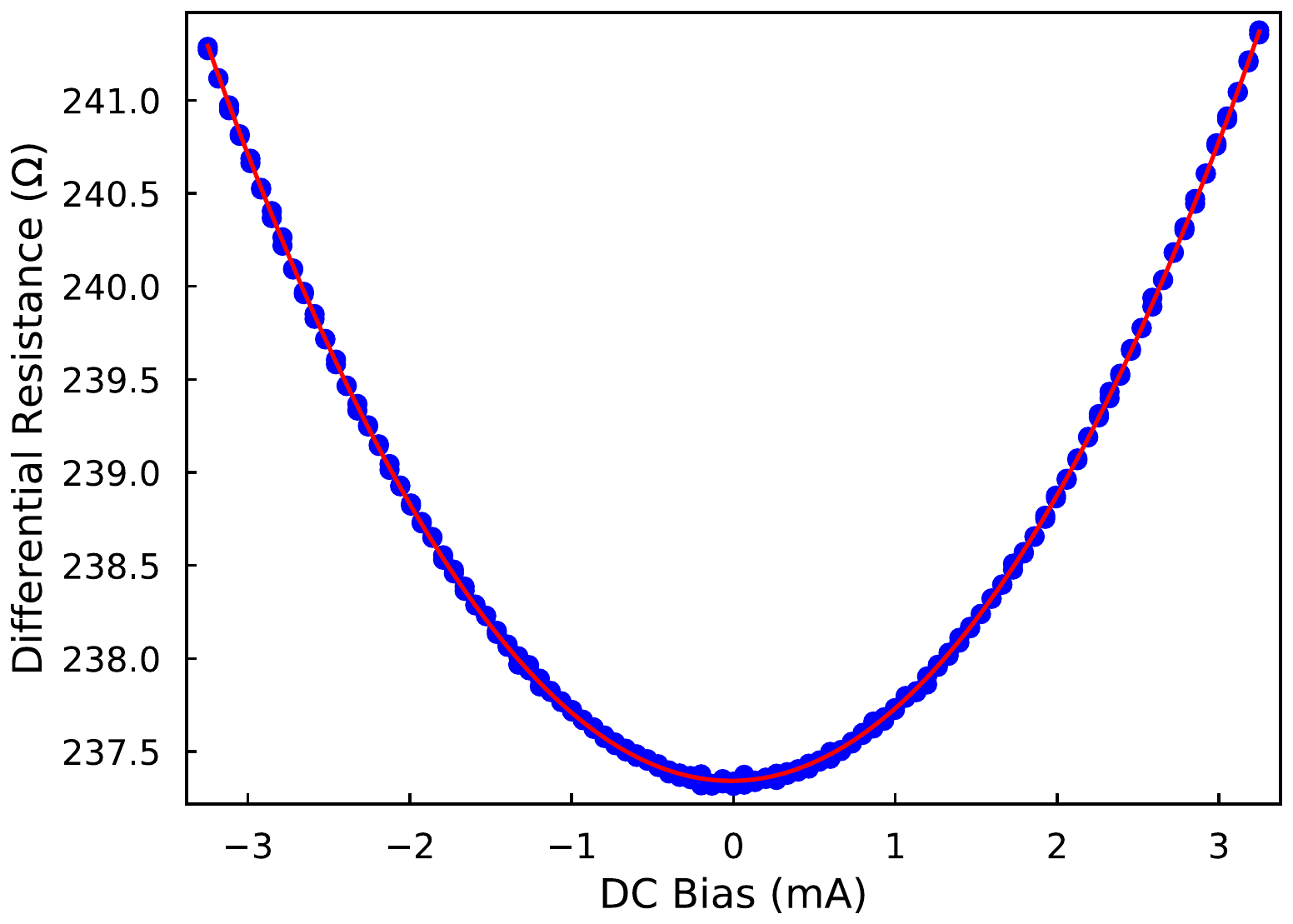}
		\par\end{centering}
	\caption{\label{fig:dVdI}Measured impulse differential resistance $R_{\text{diff,i}}$
		versus DC bias $I_{0}$ with modulation amplitude $I_{m}=81.5$ $\upmu$A,
		modulation frequency $\omega_{m}=2\pi\times265$ Hz, $N=53$ periods
		per impulse, and no RF current $(I_{1}=0)$. Solid red curve is a
		fit to the form $R_{\text{diff,s}}=\left(\partial_{I}V\right)\left(1+\chi(0)\frac{35}{64}I_{m}^{2}+3\chi(0)\left(I_{0}-I_{\text{off}}\right)^{2}\right)$,
		with fit parameters $\partial_{I}V=237.3\pm0.3\,\Omega$, $\chi(0)=5.3085\pm0.004\times10^{-4}$
		mA$^{-2}$, and offset current $I_{\text{off}}=-15\pm1$ $\upmu$A.
		Uncertainty on $\chi(0)$ and $I_{\text{off}}$ are statistical, and
		the uncertainty on $\partial_{I}V$ is systematic, due to uncertainty
		in series resistors leading to the sample. Note this measured $\partial_{I}V$
		includes the resistances of everything after the bias-T, including
		a circuit board, wire bonds, and on-chip leads / contacts, so this
		is an upper bound on the wire resistance.}
\end{figure}

In practice, the measured value of $R_{\text{diff,i}}$ does not depend
on frequency below $\sim300$ Hz (limited in our case by the low-pass
action of our bias-T), which allows us to estimate $\chi(0)$ from
the bias-dependence of $R_{\text{diff,i}}$ in the absence of RF current
($I_{1}=0$). Figure \ref{fig:dVdI} shows a measurement of $R_{\text{diff,i}}(I_{0})$,
with $\omega_{m}=2\pi\times$265 Hz. The data is well-fit by Eq.~\ref{eq:Rdiffs}
(with a small offset bias $I_{\text{off}}=-15\pm1$ $\upmu$A),
from which we extract $\partial_{I}V=237.3\pm0.3\,\Omega$, $\chi(0)=5.3085\pm0.004\times10^{-4}$
mA$^{-2}$. The small offset current is likely due to a combination
of non-ideal electronics and Peltier effects associated with asymmetric
contacts to our nanowire on chip. In the present experiment, this
offset leads to at most a few-percent error in our calibration of
$I_{1}$ (discussed below), along with a small, smoothly-varying background
signal in our ferromagnetic resonance measurements of up to $\sim$5
$\upmu$V at low frequencies. The associated correction, which
does not affect the central conclusions of the present work, will
be the subject future work.

\subsubsection{Maximum change in wire temperature}\label{subsec:joule-maximum-temperature}

In the presence of only DC bias, Eq.~\ref{eq:V} simplifies to
\begin{align}
\langle V\rangle	\rightarrow\left(\partial_{I}V\right)\left(I_{0}\right)\left(1+\chi(0)I_{0}^{2}\right),
\end{align}
such that the DC resistance
\begin{align}
\frac{\langle V\rangle}{I_{0}}	&=\left(\partial_{I}V\right)+\left(\partial_{I}V\right)\chi(0)I_{0}^{2}\\
&=R_{0}+R_{0}\alpha\Delta T,
\end{align}
where $\alpha$ is the temperature coefficient of resistance. From this, we identify 
\begin{align}
R_{0}&=\partial_{I}V
\\
\Delta T &=\frac{\chi(0)I_{0}^{2}}{\alpha}
\end{align}
Using the above fit values and using the Pt coefficient $\alpha\gtrsim 0.003$ K$^-1$ as a lower bound, we place an upper bound on the maximum Joule heating temperature change $\Delta T \lesssim 5$ K for our bias range ($I_0<5$~mA).

\subsubsection{Estimating RF current from rectified 	voltage}\label{subsec:Estimating-RF-current}

\begin{figure}[h!]
	\begin{centering}
		\includegraphics[width=9cm]{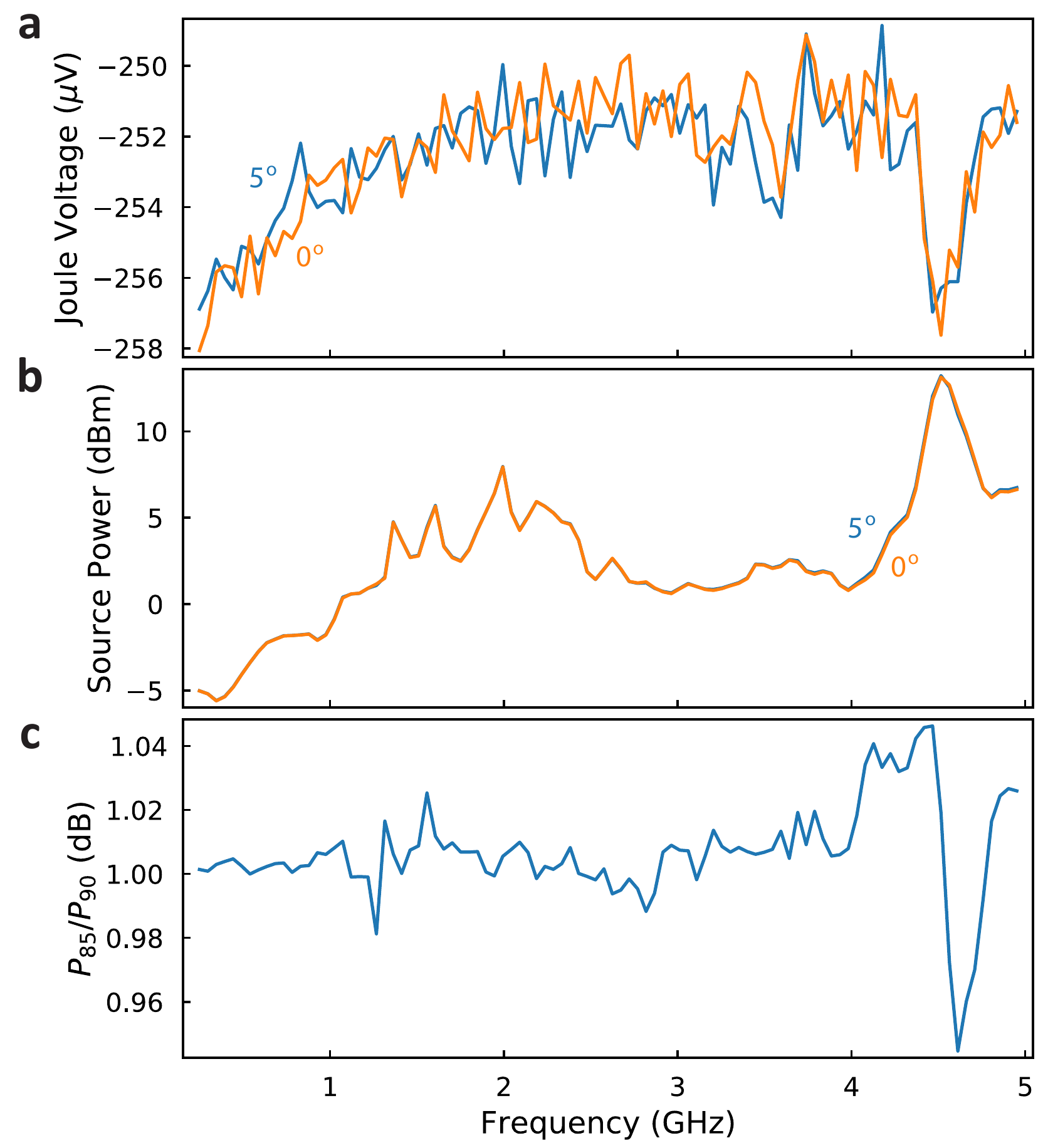}
		\par\end{centering}
	\caption{\label{fig:VmixThermal}Joule heating spectra. (a) ``Typical'' rectified
		voltage $\langle V\rangle$ versus RF frequency for bias $I_{0}=-2.04$
		mA, \emph{nominal} power -5 dBm ($\sim0.5$ mA into the nanowire),
		and field $\mu_0 H_{0}=$20.6 mT applied a in-plane angle $\theta=5^{\circ}$
		(orange) and $0^{\circ}$ (blue) from the wire axis and 35$^{\circ}$
		out of plane. The (heavily damped) magnetic resonance near 2.4 GHz
		should only appear in the off-axis ($\theta=5^{\circ}$) case, and
		evidently represents at most a sub-percent contribution to this signal.
		(b) Frequency-dependent output power used to compensate for losses
		in the electronic components between the RF source and nanowire. (c)
		Ratio of the two curves in (b), showing at most a few-percent difference
		between the compensation spectra.}
\end{figure}

Knowing $\chi(0)$, we can, in principle, use the measured rectified
voltage (Eq.~\ref{eq:V} rewritten) 
\begin{align}
\langle V\rangle & =I_{0}\left(\partial_{I}V\right)\left(1+\chi(0)I_{0}^{2}+\frac{1}{2}\left[\chi(0)+2\chi(\omega)\right]I_{1}^{2}\right)
\end{align}
to estimate the RF current $I_{1}$.
Figure \ref{fig:VmixThermal}(a) shows a ``typical'' spectrum of just
the RF-induced rectified voltage
\begin{align}
\Delta\langle V\rangle & =\langle V\rangle-\left[\langle V\rangle\right]_{I_{1}=0}\nonumber \\
& =\frac{1}{2}\left(\partial_{I}V\right)\left[\chi(0)+2\chi(\omega)\right]I_{0}I_{1}^{2}\label{eq:D<V>}
\end{align}
as recorded with a chopped lock-in technique at 265 Hz \cite{Sankey2006Spin}\footnote{The only difference from Ref.~\cite{Sankey2006Spin} is that we employ a Wheatstone bridge on the low-frequency port of our bias-T to eliminate common-mode noise from our current source, and use the ``single-shot'' readout discussed in Sec.~\ref{subsec:Single-shot-lock-in-measurement}.}, with a bias $I_{0}=-2.04$~mA, a nominal power -5 dBm ($\sim0.5$~mA into the nanowire), and field $\mu_0H_{0}=$20.6 mT applied along an in-plane angle $\theta=5^{\circ}$ (orange) and $0^{\circ}$ (blue) from the wire's long axis and 35$^{\circ}$ out of plane. Under these conditions, the magnetization dynamics are heavily damped, and the magnetization is saturated approximately along the in-plane short axis $(\hat{y})$ of the nanowire (canted a few degrees out of plane). 

Importantly, this data was recorded \emph{after} adjusting the output power at each frequency (Fig.~\ref{fig:VmixThermal}(b)) to ensure a frequency-independent $\Delta\langle V\rangle$ at $I_{0}=-4.08$~mA, which allows us to roughly compensate for the strongly frequency-dependent input coupling of our microwave circuitry. Figure \ref{fig:VmixThermal}(c) shows the ratio of the powers in b, highlighting that the damped ferromagnetic resonance feature, which should appear near 2.4 GHz in the $\theta=5^{\circ}$ spectrum, represents at most a $\sim$1\% correction. The high-frequency peak in Fig.~\ref{fig:VmixThermal}(a) is a noise feature owing to poor coupling near 4.7 GHz. The low-frequency tail is a bias-independent feature that we suspect arises from electronics nonidealities and / or a small Peltier effect. As such, we subtract it from all spectra in the main text. Its presence or absence from the data does not quantitatively or qualitatively affect the conclusions of this work.

Having no reliable means of measuring the loss spectrum of our circuit
board, wire bonds (as connected to the sample), and microfabricated
leads, we cannot use the measurement in Fig.~\ref{fig:VmixThermal}(a)
to estimate $\chi(\omega)$ directly. We can, however, still solve
Eq.~\ref{eq:D<V>} for the RF amplitude
\begin{equation}
I_{1}=\sqrt{\frac{2\Delta\langle V\rangle}{I_{0}\left(\partial_{I}V\right)\chi(0)\left(1+2\frac{\chi(\omega)}{\chi(0)}\right)}},\label{eq:I1}
\end{equation}
and, since we expect $0<\chi(\omega)<\chi(0)$ (i.e., $\chi$ is largest
at $\omega=0$), we can constrain the RF current $I_{1}$ to the range
\begin{equation}
\sqrt{\frac{2\Delta\langle V\rangle}{3I_{0}\left(\partial_{I}V\right)\chi(0)}}<I_{1}<\sqrt{\frac{2\Delta\langle V\rangle}{I_{0}\left(\partial_{I}V\right)\chi(0)}},\label{eq:I1-bounds}
\end{equation}
representing a maximum systematic error of $\pm37\%$. However, the
frequency dependence of $\chi(\omega)$ \emph{will} impose a \emph{frequency-dependent}
systematic error, which can in principle skew observed ferromagnetic
resonance curves. In the following section, we improve upon this estimate
by simulating the thermal response of our wire on a diamond substrate.

\subsection{Modeling the thermal transfer
	function for our wire}\label{subsec:Simulated-thermal-transfer}

The goal of this section is to simulate the thermal response of our
nanowire so that we may more accurately estimate how the Joule-rectified
voltage depends on the RF current and DC bias. Inspecting Eq.~\ref{eq:I1},
the only quantity of which we do \emph{not} have an independent measure
is the ratio $\chi(\omega)/\chi(0)$,\footnote{Recall the Fig.~\ref{fig:VmixThermal} provides the Joule-rectified
	voltage $\Delta\langle V\rangle$, $I_{0}$ is measured with a series
	resistor, and $\left(\partial_{I}V\right)\chi(0)$ can be estimated
	as in Fig.~\ref{fig:dVdI}.} which the following sections address.

\subsubsection{Parallel resistor model}\label{subsec:Parallel-resistor-model}

To model the thermal response of our multilayer nanowire to an oscillatory
current, we first must estimate the current carried by each layer.
We model the Pt and Py layers of the wire as two resistors connected
in parallel, with values $R_{\text{Pt}}$ and $R_{\text{Py}}$, such
that our \emph{total }current through the wire
\begin{equation}
I=I_{0}+I_{1}\cos\omega t,
\end{equation}
is divided into layer currents
\begin{align}
I_{j} & =\frac{R}{R_{j}}\left(I_{0}+I_{1}\cos\omega t\right)
\end{align}
for $j\in\{\text{Pt},\text{Py}\}$, where
\begin{equation}
R\equiv\frac{R_{\text{Py}}R_{\text{Pt}}}{R_{\text{Py}}+R_{\text{Pt}}}
\end{equation}
is the total wire resistance. The dissipated power in each layer is
then
\begin{align}
P_{j} & =\frac{R^{2}}{R_{j}}\left(I_{0}^{2}+2I_{0}I_{1}\cos\omega t+I_{1}^{2}\cos^{2}\omega t\right).
\end{align}
and the first harmonic has amplitude
\begin{equation}
P_{j1}=\frac{2I_{0}I_{1}R^{2}}{R_{j}},
\end{equation}
Importantly, $P_{j1}\propto I_{1}$, so the ratio of first-harmonic
amplitudes is 
\begin{equation}
\frac{P_{\text{Py}1}}{P_{\text{Pt}1}}=\frac{R_{\text{Pt}}}{R_{\text{Py}}}.\label{eq:power-ratio}
\end{equation}

With this oscillatory drive, we expect a steady state solution to
cause a (time dependent) temperature change $\Delta T_{j}$, such
that the layer resistances
\begin{align}
R_{j} & =R_{j0}\left(1+\alpha_{j}\Delta T_{j}\right)
\end{align}
where $I_{j}$ is the layer current, $R_{j0}$ is the zero-heat layer
resistance, and $\alpha_{j}$ is the layer's temperature coefficient
of resistance. The resistance of the combined nanowire is then
\begin{align}
R & =R_{0}\frac{\left(1+\alpha_{\text{Pt}}\Delta T_{\text{Pt}}\right)\left(1+\alpha_{\text{Py}}\Delta T_{\text{Py}}\right)}{1+\frac{R_{0}}{R_{\text{Py}}}\alpha_{\text{Pt}}\Delta T_{\text{Pt}}+\frac{R_{0}}{R_{\text{Pt}}}\alpha_{\text{Py}}\Delta T_{\text{Py}}}
\end{align}
with zero-heat total wire resistance
\begin{equation}
R_{0}\equiv\frac{R_{\text{Py0}}R_{\text{Pt0}}}{R_{\text{Py0}}+R_{\text{Pt0}}}.
\end{equation}
Assuming the resistance changes due to Joule heating are a small fraction
of the zero-heating values ($\alpha_{j}\Delta T_{j}\ll1$; see Fig.~\ref{fig:dVdI}),
\begin{equation}
R\approx R_{0}+\frac{R_{0}^{2}}{R_{\text{Pt}}}\alpha_{\text{Pt}}\Delta T_{\text{Pt}}+\frac{R_{0}^{2}}{R_{\text{Py}}}\alpha_{\text{Py}}\Delta T_{\text{Py}},\label{eq:R-1st-order}
\end{equation}
and the power amplitudes in each layer (keeping only terms of order
$I_{j}^{2}\sim\Delta T_{j}$) become
\begin{align}
P_{j} & \approx\frac{R_{0}^{2}}{R_{j0}}\left(I_{0}^{2}+2I_{0}I_{1}\cos\omega t+I_{1}^{2}\cos^{2}\omega t\right).
\end{align}
In the presence of these static and oscillatory powers, the layer
temperature change
\begin{align}
\Delta T_{j} & =\left(X_{j}(0)P_{j0}+P_{j1}\left(X_{j}(\omega)\cos\omega t+Y_{j}(\omega)\sin\omega t\right)+X_{j}(2\omega)\cos2\omega t\right)\\
P_{j0} & \equiv\left(I_{0}^{2}+\frac{1}{2}I_{1}^{2}\right)\frac{R_{0}^{2}}{R_{j}}\\
P_{j1} & \equiv2I_{0}I_{1}\frac{R_{0}^{2}}{R_{j}}\\
P_{j2} & \equiv\frac{1}{2}I_{1}^{2}\frac{R_{0}^{2}}{R_{j}},
\end{align}
where $X_{j}(\omega)$ and $Y_{j}(\omega)$ are the quadratures of
the transfer function for each layer. The instantaneous \emph{total
}voltage is then
\begin{align}
V & =IR\\
& =R_{0}\left(I_{0}+I_{1}\cos\omega t\right)\left(1+\sum_{j}\frac{R_{0}}{R_{j}}\alpha_{j}X_{j}(0)P_{j0}\right)\\
& +R_{0}\left(I_{0}+I_{1}\cos\omega t\right)\left(\sum_{j}\frac{R_{0}}{R_{j}}\alpha_{j}P_{j1}\left(X_{j}(\omega)\cos\omega t+Y_{j}(\omega)\sin\omega t\right)\right)\\
& +R_{0}\left(I_{0}+I_{1}\cos\omega t\right)\left(\sum_{j}\frac{R_{0}}{R_{j}}\alpha_{j}X_{j}(2\omega)\cos2\omega t\right),
\end{align}
and the time-averaged voltage can be written as
\begin{equation}
\langle V\rangle=I_{0}R_{0}\left(1+\chi(0)I_{0}^{2}+\frac{1}{2}\left[\chi(0)+2\chi(\omega)\right]I_{1}^{2}\right)
\end{equation}
with the \emph{total} in-phase transfer function
\begin{align}
\chi(\omega) & \equiv\sum_{j}\frac{R_{0}^{3}}{R_{j}^{2}}\alpha_{j}X_{j}(\omega)\nonumber \\
& =\frac{R_{0}^{3}}{R_{\text{Pt}}^{2}}\alpha_{\text{Pt}}X_{\text{Pt}}(\omega)+\frac{R_{0}^{3}}{R_{\text{Py}}^{2}}\alpha_{\text{Py}}X_{\text{Py}}(\omega).\label{eq:chi-2-layer}
\end{align}
This formula is identical to that of the single-resistor case (Eq.~\ref{eq:V}
of Sec.~\ref{subsec:Derivation-of-Joule}, identifying $R_{0}=\partial_{I}V$),
except that $\chi(\omega)$ is now a weighted average of the layers'
individual responses. Not surprisingly, the weighting factors scale
as $\alpha_{j}$, and increase as the layer resistance $R_{j}$ is
reduced (when a larger fraction of the current flows through layer
$j$). Additionally, for the case of a single layer (of resistance
$R_{0}$, thermal coefficient $\alpha$, and transfer function $X$),
this expression simplifies to $\chi(\omega)\rightarrow R\alpha X(\omega)$
and we can identify $\partial_{T}\partial_{I}V=R_{0}\alpha$ from
Eq.~\ref{eq:V}, as expected. 

As mentioned above (see also Eq.~\ref{eq:I1}), we are interested
in the ratio
\begin{align}
\frac{\chi(\omega)}{\chi(0)} & =\left(\frac{X_{\text{Pt}}(\omega)}{X_{\text{Pt}}(0)}\right)\frac{1+\eta X_{\text{Py}}(\omega)/X_{\text{Pt}}(\omega)}{1+\eta X_{\text{Py}}(0)/X_{\text{Pt}}(0)}\label{eq:chiw/chi0}\\
\eta & =\frac{R_{\text{Pt}}^{2}\alpha_{\text{Py}}}{R_{\text{Py}}^{2}\alpha_{\text{Pt}}}.\nonumber 
\end{align}
Assuming $R_{\text{Pt}}/R_{\text{Py}}=0.325$ \cite{Duan2014Spin}
and $\alpha_{\text{Py}}/\text{\ensuremath{\alpha}}_{\text{Pt}}\sim1.33$
for our system, $\eta\sim0.15$. As we will show in the following
section, $X_{\text{Py}}/X_{\text{Pt}}<1$ over the frequency range
of interest, so we can make a further approximation
\begin{equation}
\frac{\chi(\omega)}{\chi(0)}\approx\left(\frac{X_{\text{Pt}}(\omega)}{X_{\text{Pt}}(0)}\right)\left(1+\eta\left[\frac{X_{\text{Py}}(\omega)}{X_{\text{Pt}}(\omega)}-\frac{X_{\text{Py}}(0)}{X_{\text{Pt}}(0)}\right]\right)\label{eq:chiw/chi0-approximate}
\end{equation}
which illustrates the correction due to the presence of the Py layer
is small (as discussed below, it should be $\lesssim3\%$). Nonetheless,
we simulate both layers because there is not much additional overhead.

\begin{figure}[h!]
	\noindent \begin{centering}
		\includegraphics[width=8cm]{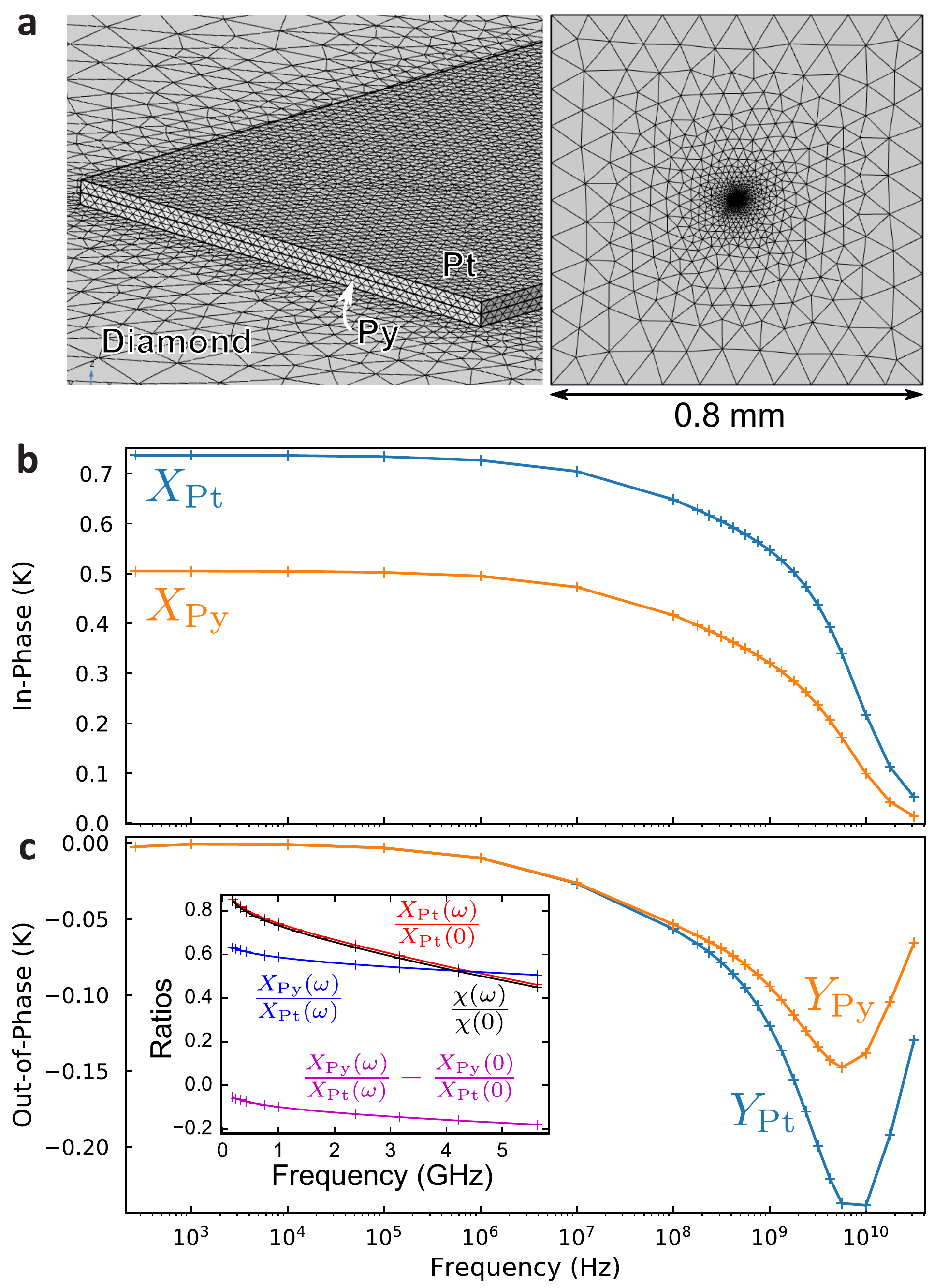}
		\par\end{centering}
	\caption{\label{fig:Simulated-thermal-response}Simulated thermal response
		of Py/Pt nanowire on diamond, assuming a wire length $l$=5.2 $\upmu$m,
		width $w$=417 nm, Pt and Py thicknesses $h$=10 nm, resistivities
		$\rho_{\text{Pt}}=21.9\,\Omega$-cm and $\rho_{\text{Py}}=65.2\,\Omega$-cm.
		(a) Simulation volume for the low-frequency response (left) and nanowire
		mesh (right). (b) In-phase temperature oscillation amplitude at the
		drive frequency for the Pt (blue) and Py (orange) layer, for a drive
		power amplitude $P_{\text{Pt}1}=10^{17}$ W/m$^{3}$ and $P_{\text{Py1}}=0.336\times10^{17}$
		W/m$^{3}$ in the Py layer. (b) Out-of-phase oscillation amplitude,
		showing steady-state behavior at low frequencies (where the thermal
		penetration depth in the diamond is much larger than the wire) and
		the onset of a lagged response at higher frequencies (where the heat
		is confined to very near the wire). Inset shows $\chi(\omega)/\chi(0)$
		(black) as well as $X_{\text{Pt}}(\omega)/X_{\text{Pt}}(0)$ (red),
		$X_{\text{Py}}(\omega)/X_{\text{Pt}}(\omega)$ (blue), $\frac{X_{\text{Py}}(\omega)}{X_{\text{Pt}}(\omega)}-\frac{X_{\text{Py}}(0)}{X_{\text{Pt}}(0)}$
		(magenta) from Eqs.~\ref{eq:chiw/chi0} and \ref{eq:chiw/chi0-approximate}.}
\end{figure}

\subsubsection{Finite-element simulation}

We perform finite-element simulations (COMSOL) of the nanowire's active
region (between the contacts, where the current is concentrated) and
the diamond substrate. Figure \ref{fig:Simulated-thermal-response}(a)
shows the simulated geometry, comprising a $L\times W\times H$ diamond
substrate (dimensions varied with frequency as discussed below), upon
which a 10 nm Py (bottom) / 10 nm Pt (top) nanowire spanning 5.2 $\times$
0.417 $\upmu$m$^{2}$ is positioned at the center. We assume
the heat is primarily dissipated in the constriction (the nanowire),
and that the large-area connection to diamond serves as the dominant
heat sink in this system; as such we do not bother to include the leads in this
study. Including a ``standard'' convective heat loss of $\sim$25
W/m$^{2}$K boundary condition on all free surfaces does not significantly alter the results, nor does reducing the mesh density along each axis
by a factor of 2.

As per Eq.~\ref{eq:power-ratio}, we introduce an oscillatory power
of amplitude $P_{\text{Pt1}}$=10$^{17}$ W/m$^{3}$ and $P_{\text{Py1}}=\left(R_{\text{Py}}/R_{\text{Pt}}\right)P_{\text{Pt1}}$
and frequency $\omega$ to the Pt and Py layers. We let the simulation
run until it converges to a steady state, then extract the in-phase
($X_{j}$, Fig.~\ref{fig:Simulated-thermal-response}(b)) and out-of-phase ($Y_{j}$,
Fig.~\ref{fig:Simulated-thermal-response}(c)) amplitudes from the time-dependent
temperatures $T_{j}(t)$.\footnote{Note this single-frequency method is significantly more efficient
	than performing an impulse response when covering this many decades
	of frequency.} 

At lower frequencies, the temperature change penetrates further into
the diamond, and a larger diamond volume is required for the results
to converge. At the lowest frequency simulated ($\omega=2\pi\times$265
Hz), for example, we employed the 0.8 mm $\times$ 0.8 mm $\times$
0.4 mm volume shown in Fig.~\ref{fig:Simulated-thermal-response}(a), but a small
deviation from adiabatic response is still evidenced by a sub-percent
out-of-phase response due to integrator-like behavior as heat reaches
the boundaries of the diamond. This phase lag returns at
higher frequencies ($\omega>2\pi\times10^{6}$ Hz) as the heat ceases
to efficiently escape the nanowire. 

\begin{figure}[h!]
	\begin{centering}
		\includegraphics[width=8cm]{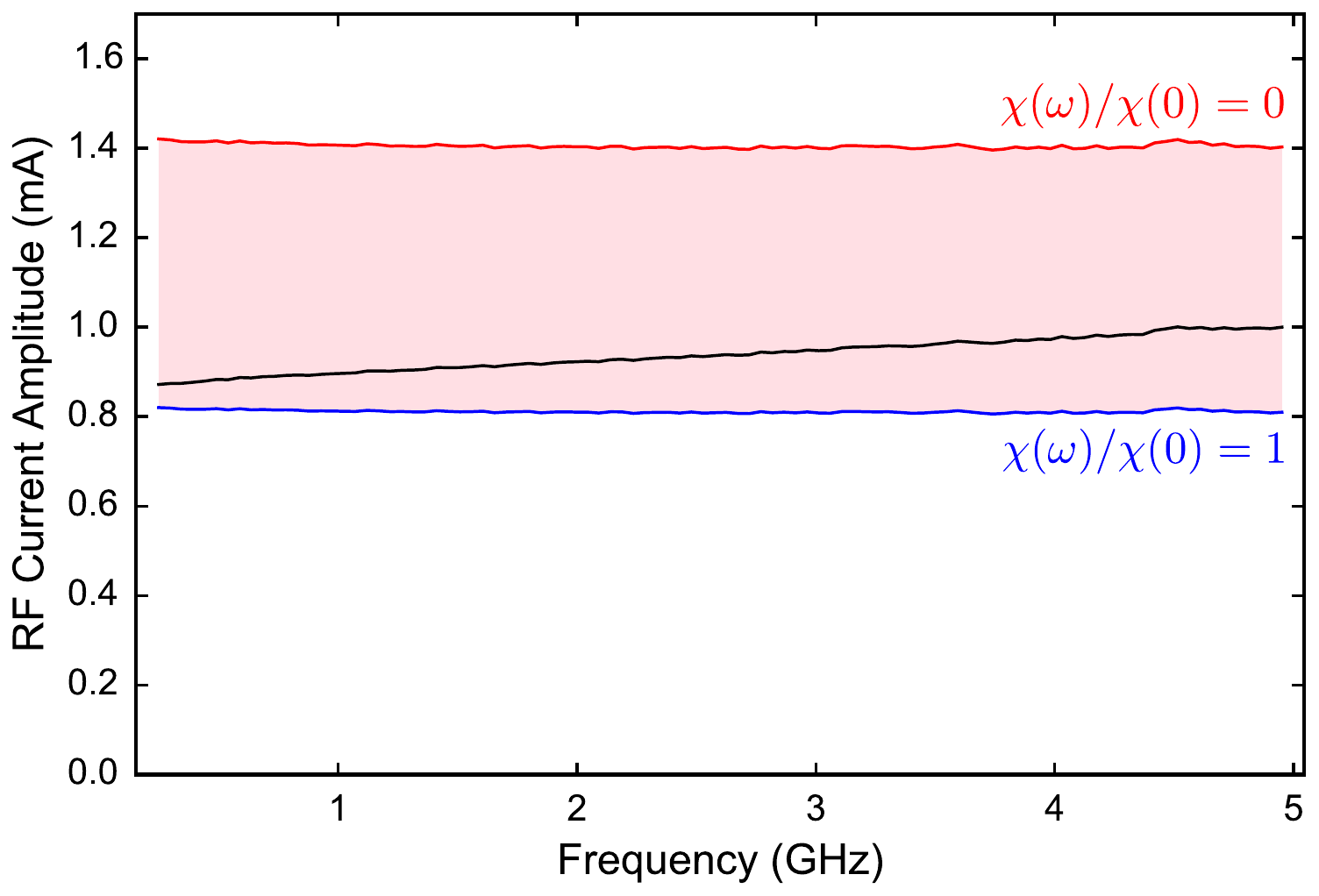}
		\par\end{centering}
	\caption{\label{fig:calibration}RF current amplitude $I_{1}$ calibrated from
		the data in Figs.~\ref{fig:dVdI}-\ref{fig:VmixThermal} (Fig.~2
		of the main text) and the simulated thermal response in Fig.~\ref{fig:Simulated-thermal-response}
		(black). Also shown is the model-independent calibration range (shaded
		region) bounded by the case $\chi(\omega)/\chi(0)=0$ (red) and $\chi(\omega)/\chi(0)=1$
		(blue).}
\end{figure}

The inset of Fig.~\ref{fig:Simulated-thermal-response}(c) shows the
quantities of interest in Eqs.~\ref{eq:chiw/chi0}-\ref{eq:chiw/chi0-approximate},
as well as $\chi(\omega)/\chi(0)$ for the \emph{total} wire assuming
$\eta=0.15$. We can now plug this into our expression for the RF
current amplitude (Eq.~\ref{eq:I1} rewritten)
\begin{equation}
I_{1}=\sqrt{\frac{2\Delta\langle V\rangle}{I_{0}\left(\partial_{I}V\right)\chi(0)\left(1+2\frac{\chi(\omega)}{\chi(0)}\right)}}
\end{equation}
to convert the observed rectified voltage $\Delta\langle V\rangle$ in Fig.~\ref{fig:VmixThermal}(a)
(at $I_{0}=-2.04$ mA) to RF current, using the values $\partial_{I}V=237.3\pm0.3\,\Omega$,
$\chi(0)=5.3085\pm0.004\times10^{-4}$ mA$^{-2}$ from the fit in Fig.~\ref{fig:dVdI}. Figure \ref{fig:calibration} shows the frequency dependence of $I_{1}$ (black curve) along with the absolute calibration range of Eq.~\ref{eq:I1-bounds} (shaded region), highlighting the importance of the frequency-dependent thermal response in our system. If we ignored $\chi(\omega)$ as in Ref.~\cite{Tshitoyan2015Electrical} (e.g.) we would overestimate $I_{1}$ by up to 65\%, and miss a systematic variation of $\sim10\%$ over the probed frequency range. We suspect this correction is smaller in more thermally-isolated systems such as wires on oxide or spin valves, though we note that the thermal transfer function falls off very gradually at high frequencies.

\pagebreak
\section{Macrospin model}\label{sec:macrospin}

In this section we develop a useful toy model in which the magnetization of the Py layer is assumed to be spatially uniform. Section \ref{subsec:macrospin-equations} presents the equations of motion, Sec.~\ref{subsec:macrospin-small-angle-precession} derives the natural frequency for small-angle precession about equilibrium, and Sec.~\ref{subsec:macrospin-linear-susceptibility} derives the magnetization's linear susceptibility to general oscillatory torques. Finally, we perform simulations to roughly describe the parametrically driven, large-amplitude dynamics in Figs.~3-4 of the main text in Sec.~\ref{subsec:macrospin-parametric}.

\subsection{Equation of motion}\label{subsec:macrospin-equations}

We employ the Landau-Lifshitz-Gilbert equation in the low-damping limit, with the spin Hall torque from the Pt layer \cite{Liu2011Spin}:
\begin{align}
\partial_{t}\hat{m} & =\gamma_{0}\mu_{0}\vec{H}_{\text{eff}}\times\hat{m}+\alpha\hat{m}\times\partial_{t}\hat{m}+\frac{g\mu_{B}\eta\theta_{\text{SH}}J_{\text{Pt}}}{2et_{Py}M_{s}}\hat{m}\times\left(\hat{m}\times\hat{y}\right)\nonumber \\
& \approx\gamma_{0}\mu_{0}\left(\vec{H}_{\text{eff}}\times\hat{m}+\alpha\hat{m}\times\left(\vec{H}_{\text{eff}}\times\hat{m}\right)\right)+\frac{g\mu_{B}\eta\theta_{\text{SH}}J_{\text{Pt}}}{2et_{\text{Py}}M_{\text{s}}}\hat{m}\times\left(\hat{m}\times\hat{y}\right).\label{eq:dmdt}
\end{align}
Here, $\hat{m}=m_{x}\hat{x}+m_{y}\hat{y}+m_{z}\hat{z}$ is a unit
vector describing the orientation of the magnetization, $\gamma_{0}$ is the magnitude of the gyromagnetic ratio, $\mu_{0}$ is the
magnetic permeability of free space, $\vec{H}_{\text{eff}}$ is the
effective field (discussed below), $\alpha$ is the damping parameter
(approximately equal to the Gilbert damping for weak damping and
spin transfer torques), $\mu_{B}$ is the
Bohr magneton, $\eta$ is the fraction of incident spins that are
absorbed by the Py layer, $\theta_{\text{SH}}$ is the spin Hall angle of Pt, $J_{\text{Pt}}$ is the charge current density in the Pt layer (determined by the parallel resistor model of Sec.~\ref{subsec:Parallel-resistor-model}), $e$ is the electron charge, $t_{\text{Py}}$ is the thickness
of the Py layer, and $M_{\text{s}}$ is its effective saturation magnetization.
The effective field
\begin{equation}
\vec{H}_{\text{eff}}=\vec{H}_{0}+\vec{H}_{\text{an}}+\vec{H}_{I},
\end{equation}
where $\vec{H}_{0}=H_{x}\hat{x}+H_{y}\hat{y}+H_{z}\hat{z}$ is the
applied field, the shape anisotropy field
\begin{align}
\vec{H}_{\text{an}} & =-M_{\text{s}}\left(N_{yy}-N_{xx}\right)m_{y}\hat{y}-M_{\text{s}}\left(N_{zz}-N_{xx}\right)m_{z}\hat{z}\\
& \equiv-H_{yx}m_{y}\hat{y}-H_{zx}m_{z}\hat{z},
\end{align}
with $N_{ij}$ being the elements of
a demagnetization tensor (assumed diagonal for simplicity, with
$N_{xx}+N_{yy}+N_{zz}=1$),
and 
\begin{equation}
\vec{H}_{I}=-aI(t)\hat{y}
\end{equation}
is the spatially averaged field generated by the instantaneous current
$I(t)$ flowing through the wire, with proportionality constant $a$. For our long wire (aligned along $x$), we assume $N_{xx}\ll N_{yy}\ll N_{zz}\sim1$,
such that $H_{yx}+H_{zx}\approx M_{\text{s}}$ is the effective magnetization. 

\subsection{Small-angle precession frequency}\label{subsec:macrospin-small-angle-precession}

Our first goal is to estimate the resonant frequency for for small-angle
precession when the applied field is sufficient to saturate $m_{y}$.
To do so, we ignore dissipation and current in Eq.~\ref{eq:dmdt},
and include an applied field with $H_x=0$ (as in the experiment):
\begin{align}
\partial_{t}\hat{m} & =\gamma_{0}\mu_{0}\vec{H}_{\text{eff}}\times\hat{m}\nonumber \\
\frac{\partial_{t}\hat{m}}{\gamma_{0}\mu_{0}} & =\left(\left(H_{y}-H_{yx}m_{y}\right)\hat{y}+\left(H_{z}-H_{zx}m_{z}\right)\hat{z}\right)\times\hat{m}\nonumber \\
& =\left(\begin{array}{c}
\left(H_{y}-H_{yx}m_{y}\right)m_{z}-\left(H_{z}-H_{zx}m_{z}\right)m_{y}\\
\left(H_{z}-H_{zx}m_{z}\right)m_{x}\\
-\left(H_{y}-H_{yx}m_{y}\right)m_{x}
\end{array}\right).\label{eq:dmdt-field-only}
\end{align}
We find the equilibrium values $m_{x0}$ and $m_{y0}$ of $m_{x}$
and $m_{y}$ by setting $\partial_{t}\hat{m}=0$, which gives three
equations and three unknowns. Assuming $H_y$ is large enough that $m_{x}=0$
leaves only the first equation, which can be written in terms of
$m_{z0}$ as

\begin{align}
\frac{H_{y}}{\sqrt{1-m_{z0}^{2}}} & =\frac{H_{z}}{m_{z0}}+H_{yx}-H_{zx}.\label{eq:mz0}
\end{align}
We note that $H_z\sim 0.5$ T is required to saturate
the magnetization along $\hat{z}$ meaning our $\sim10$ mT fields
will only slightly raise the magnetization out of the plane. As such, we assume
$m_{z0}\ll1$ so that $m_{y0}=\sqrt{1-m_{z}^{2}}\approx1-\frac{1}{2}m_{z0}^{2}$.
To first order in $m_{z0}$, this simplifies to 
\begin{align}
m_{z0} & \approx\frac{H_{z}}{H_{y}+H_{zx}-H_{yx}}.
\end{align}
For our system's effective saturation fields $\mu_{0}H_{yx}=7.57$ mT and $\mu_{0}H_{zx}=517$ (see Sec.~\ref{Sec:FMRfit}), a 35-mT field applied along the NV axis (35$^{\circ}$
out of plane), $m_{z0}\approx0.04$ (2.3$^{\circ}$ out of plane). To find the natural precession frequency $\nu_{\text{FMR}}$, we therefore apply
the limit $m_{x},m_{z}\ll1$ and $m_{y}\approx1$, to Eq.~\ref{eq:dmdt-field-only},
which yields coupled differential equations for $m_{x}$ and $m_{z}$:
\begin{align}
\frac{\partial_{t}m_{x}}{\gamma_{0}\mu_{0}} & \approx\left(H_{y}+H_{zy}\right)m_{z}-H_{z}\\
\frac{\partial_{t}m_{z}}{\gamma_{0}\mu_{0}} & \approx-\left(H_{y}-H_{yx}\right)m_{x},
\end{align}
with $H_{zy}=H_{zx}-H_{yx}$. Using the trial solution 
\begin{align}
m_{x} & =X_{0}\cos(2\pi\nu_{\text{FMR}}t)\\
m_{z} & =m_{z0}-Z_{0}\sin(2\pi\nu_{\text{FMR}}t)
\end{align}
with real-valued amplitudes $X_{0}$ and $Z_{0}$ yields 
\begin{align}
2\pi\nu_{\text{FMR}}X_{0} & \approx\gamma_{0}\mu_{0}\left(H_{y}+H_{zy}\right)Z_{0}\\
2\pi\nu_{\text{FMR}}Z_{0} & \approx\gamma_{0}\mu_{0}\left(H_{y}-H_{yx}\right)X_{0},
\end{align}
from which the amplitude ratio
\begin{equation}
\frac{X_{0}}{Z_{0}}=\sqrt{\frac{H_{y}+H_{zy}}{H_{y}-H_{yx}}}
\end{equation}
and resonant frequency
\begin{equation}\label{eq:macrospin-FMR-frequency}
\nu_{\text{FMR}}=\frac{\gamma_{0}\mu_{0}}{2\pi}\sqrt{\left(H_{y}-H_{yx}\right)\left(H_{y}+H_{zx}-H_{yx}\right)}.
\end{equation}
Importantly, this is the same (Kittel) formula one would arrive at with a purely in-plane
field, which makes sense
in this small-$m_{z}$ limit. Also, even for our maximal
in-plane field $\mu_{0}H_{y}=33$ mT, $X_{0}/Z_{0}\sim 4.7$, and this
ratio increases at lower fields, diverging at $H_{y}=H_{yx}$, as
expected. As such, the $m_z$-generated stray field power at the NV (i.e., the quantity responsible for the spin relaxation
rates) is at least $(X_{0}/Z_{0})^{2}\sim22$ times smaller than that of
$m_x$ (and much smaller at the fields of interest).

\subsection{Linear susceptibility for magnetization approximately along the in-plane hard axis}\label{subsec:macrospin-linear-susceptibility}

We now derive the susceptibility of $m_{x}$
to an oscillatory drive at frequency $\nu$. Since we know the magnetization
is saturated along $\hat{y}$ to good approximation (and behaves as though
$H_{z}=0$ for our parameter range; see Sec.~\ref{subsec:macrospin-small-angle-precession}),
we consider an applied field $\vec{H}_{0}\parallel\hat{y}$ for simplicity, and include
a general infinitesimal drive torque
\begin{equation}
\partial_{t}\hat{m}=\gamma_{0}\mu_{0}\left(\delta_{x}\hat{x}+\delta_{z}\hat{z}\right)e^{i2\pi\nu t}
\end{equation}
of amplitudes $\delta_{x}$ and $\delta_{z}$. Equation~\ref{eq:dmdt} can then be written (replacing the current-induced terms with this torque) as
\begin{eqnarray}
	\frac{\partial_{t}\hat{m}}{\gamma_{0}\mu_{0}} & = & \left(\vec{H}_{0}+\vec{H}_{\text{an}}-\alpha\left(\vec{H}_{0}+\vec{H}_{\text{an}}\right)\times\hat{m}\right)\times\hat{m}+\left(\delta_{x}\hat{x}+\delta_{z}\hat{z}\right)e^{i2\pi\nu t}.
\end{eqnarray}
In the limit $m_{x},m_{z}\ll1$ and $m_{y}\approx1$ to first order, so this becomes
\begin{eqnarray}
	\frac{\partial_{t}\hat{m}}{\gamma_{0}\mu_{0}} & \approx & \left(\begin{array}{c}
		-\alpha\left(H_{y}+H_{zy}\right)m_{z}\\
		H_{y}-H_{yx}\\
		-H_{zx}m_{z}+\alpha\left(H_{y}-H_{yx}\right)m_{x}
	\end{array}\right)\times\left(\begin{array}{c}
		m_{x}\\
		m_{y}\\
		m_{z}
	\end{array}\right)+\left(\begin{array}{c}
		\delta_{x}\\
		0\\
		\delta_{z}
	\end{array}\right)e^{i2\pi\nu t}
\end{eqnarray}
or
\begin{eqnarray}
	\frac{\partial_{t}m_{x}}{\gamma_{0}\mu_{0}} & \approx & \left(H_{y}+H_{zy}\right)m_{z}-\alpha\left(H_{y}-H_{yx}\right)m_{x}+\delta_{x}e^{i2\pi\nu t}\\
	\frac{\partial_{t}m_{z}}{\gamma_{0}\mu_{0}} & \approx & -\alpha\left(H_{y}+H_{zy}\right)m_{z}-\left(H_{y}-H_{yx}\right)m_{x}+\delta_{z}e^{i2\pi\nu t}.
\end{eqnarray}
Using the trial solution
\begin{eqnarray}
	m_{x} & = & \tilde{X}_{0}e^{i2\pi\nu t}\\
	m_{z} & = & \tilde{Z}_{0}e^{i2\pi\nu t}
\end{eqnarray}
with complex amplitudes $\tilde{X}_{0}$ and $\tilde{Z}_{0}$ yields
an in-plane steady state amplitude
\begin{eqnarray}
\tilde{X}_{0}(\nu) & \approx & \left(\frac{\gamma_{0}\mu_{0}}{2\pi}\right)\left(\frac{\nu_{zy}\delta_{z}+i\nu\delta_{x}}{\nu_{\text{FMR}}^{2}-\nu^{2}+i\nu\Delta\nu}\right)\label{eq:mx-response}
\end{eqnarray}
with 
\begin{eqnarray}
	\Delta\nu & \equiv & 2\alpha\frac{\gamma_{0}\mu_{0}}{2\pi}\left(H_{y}-H_{yx}+\frac{1}{2}H_{zx}\right)\label{eq:macrospin-fmr-linewidth}\\
	\nu_{zy} & \equiv & \frac{\gamma_{0}\mu_{0}}{2\pi}\left(H_{y}+H_{zy}\right)\\
	\nu_{yx} & \equiv & \frac{\gamma_{0}\mu_{0}}{2\pi}\left(H_{y}-H_{yx}\right)\\
	\nu_{\text{FMR}} & \equiv & \sqrt{\nu_{zy}\nu_{yx}}.
\end{eqnarray}
The linewidth nominally depends on $H_{y}$, but $H_{zx}$ is the dominant effect, and so $\Delta\nu$ should remain approximately constant over our (small) field range. Also, torques along $\hat{x}$ and $\hat{z}$ produce qualitatively different lineshapes, in principle enabling a torque vector measurement \cite{Sankey2008Measurement,Kubota2008Quantitative,Liu2011Spin}.

\subsubsection{Mixdown voltage and FMR fit function}\label{subsec:macrospin-mixdown}

If the applied field is tilted slightly toward $\hat{x}$, the equilibrium
magnetization will gain a small component $m_{x0}$. As
long as $m_{x0}\ll1$, the response $\tilde{X}_{0}(\nu)$ should not
change to first order (by symmetry\footnote{The resonant frequency $\nu_{\text{FMR}}$ increases $\propto m_{x0}^{2}$ to lowest order, since the anisotropy field maximally opposes the applied field when aligned with $\hat{y}$.}).
However such a tilt \emph{does} provide access to an in-phase anisotropic
magnetoresistance (AMR) oscillation with phase delay $\psi$, amplitude $\Delta R_\text{RF}$, and in-phase component $\Delta R_{\text{RF}}\cos\psi$
(see Sec.~\ref{sec:transport}) proportional to the real part of $\tilde{X}_{0}$,
which can be written (see Eq.~\ref{eq:mx-response})
\begin{equation}\label{eq:mx-response-real-part}
\text{Re}\left[\tilde{X}_{0}(\nu)\right]\approx\left(\frac{\gamma_{0}\mu_{0}}{2\pi}\right)\frac{\delta_{z}\nu_{zy}\left(\nu_{\text{FMR}}^{2}-\nu^{2}\right)+\delta_{x}\nu^{2}\Delta\nu}{\left(\nu_{\text{FMR}}^{2}-\nu^{2}\right)^{2}+\nu^{2}\Delta\nu^{2}}.
\end{equation}
For small-angle precession, the static change in resistance $\Delta R_{0}$
should contribute very little to the FMR signal, but would scale as
\begin{equation}
\left|\tilde{X}_{0}(\nu)\right|^{2}\propto\frac{\nu_{zy}^{2}\delta_{z}^{2}+\nu^{2}\delta_{x}^{2}}{\left(\nu_{\text{FMR}}^{2}-\nu^{2}\right)^{2}+\nu^{2}\Delta\nu^{2}},
\end{equation}
which is close in form to Re$\left[\tilde{X}_{0}(\nu)\right]$. As
such, FMR spectra are well fit by
\begin{equation}\label{eq:macrospin-FMR-fit-function}
V_{\text{MR}}(\nu)=\frac{a_{z}\left(\nu_{\text{FMR}}^{2}-\nu^{2}\right)+a_{x}\nu^{2}\Delta\nu}{\left(\nu_{\text{FMR}}^{2}-\nu^{2}\right)^{2}+\nu^{2}\Delta\nu^{2}},
\end{equation}
with free parameters $a_{z}$, $a_{x}$, $\nu_{\text{FMR}}$, and $\Delta\nu$.

\subsubsection{Expected spectrum of Brownian magnetization noise}\label{subsec:macrospin-brownian}

A thermal (Langevin) field \cite{Brown1963Thermal} is typically assumed isotropic, exerting stochastic, uncorrelated torques with a white noise spectrum in all three dimensions. If the torque power
spectral densities of each component are $S_{T}$ (units of rad$^{2}$
sec$^{-2}$ Hz$^{-1}$), then we expect the power spectral density
of $m_{x}$ to be scaled by the magnitude of the susceptibility squared:
\begin{eqnarray}
S_{m_{x}} & \propto & \left|\frac{\nu_{zy}}{\nu_{\text{FMR}}^{2}-\nu^{2}+i\nu\Delta\nu}\right|^{2}S_{T}+\left|\frac{i\nu}{\nu_{\text{FMR}}^{2}-\nu^{2}+i\nu\Delta\nu}\right|^{2}S_{T}\nonumber \\
& \propto & \frac{\nu_{zy}^{2}+\nu^{2}}{\left(\nu_{\text{FMR}}^{2}-\nu^{2}\right)^{2}+\left(\nu\Delta\nu\right)^{2}}.\label{eq:brownian-noise-spectrum}
\end{eqnarray}

\subsection{Parametrically driven, large-amplitude oscillations}\label{subsec:macrospin-parametric}

The transport technique used to measure the signals shown in Fig.~2
of the main text are most sensitive to the most spatially uniform
magnetic oscillations. To gain some qualitative intuition about
the observed large-amplitude, parametrically driven dynamics, we numerically integrate Eq.~\ref{eq:dmdt} with $\gamma_{0}=2\pi\times29.25$
GHz/T (electron g-factor
$g=2.09$ for 10-nm-thick Py \cite{Nibarger2003Variation}), $\eta\theta_{\text{SH}}=0.055$ \cite{Liu2011Spin}, layer resistivity ratio $\rho_{\text{Py}}/\rho_{\text{Pt}}=2.977$,
and $t_{\text{Py}}=10$ nm. From our fits (Sec.~\ref{Sec:FMRfit})
we use $\mu_{0}H_{yx}=7.57$ mT and $\mu_{0}H_{zx}=517$ mT and effective magnetization $\mu_{0}M_{\text{s}}=525$ mT for consistency. The proportionality constant for current-induced field $a=1.72$ mT/mA is estimated from
the required compensation field at 4.08 mA in Fig.~2 of the main
text. In order to achieve a similar threshold for parametric oscillations (i.e., not needing unreasonably large currents), we choose $\alpha=0.02$, which is approximately half the value estimated from our FMR fits. This discrepancy is under study, but we nominally think it is related to the actual device's nonuniform magnetization and / or other nonidealities such as material contamination (e.g., oxidization) and roughness.\footnote{Macrospin models are often surprisingly accurate in describing actual nanostructured systems, but their results should always be considered with caution.}

\begin{figure}
	\begin{centering}
		\includegraphics[width=10cm]{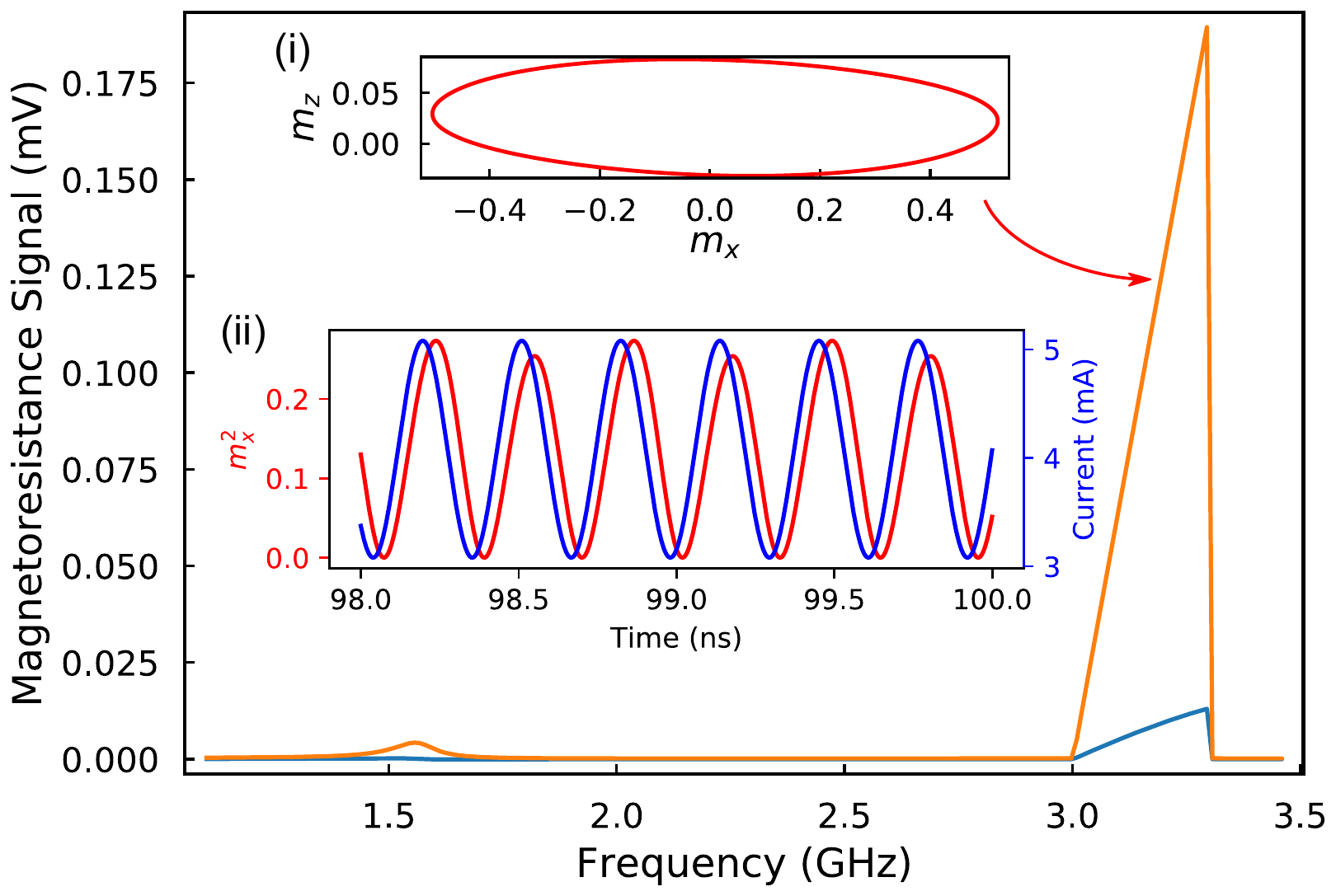}
		\par\end{centering}
	\caption{\label{fig:macrospyn-parametric}Simulated magnetoresistance signal
		$\Delta V_\text{MR}$ (orange) and the contribution $\Delta V_{\text{mix}}$
		from mixing with RF current $I_{1}$ alone (blue). Simulation parameters are listed in the main text. 
		At this value of DC bias $I_{0} = 4.08$ mA, the signal is dominated by $I_{0}\left\langle \Delta R(t)\right\rangle $,
		which has a different line shape at the fundamental frequency. (i)
		Trajectory of $\hat{m}$ at $\nu_{1}=3.1875$ GHz, where $\Delta V_\text{MR} = 120$~$\upmu$V. (ii)
		Time domain showing the relative phase of the resistance ($\propto m_{x}^{2}$)
		and current oscillations.}
\end{figure}

To mimic the red data in Fig.~2(a) of the text, we apply a field
$\mu_{0}H_{0}=23.5$ mT $\times\left(\sqrt{\frac{2}{3}}\hat{y}+\sqrt{\frac{1}{3}}\hat{z}\right)$
along the NV axis, and current
\begin{equation}
I(t)=I_{0}+I_{1}\cos(2\pi\nu_{1}t)
\end{equation}
with $I_{0}=4.08$ mA, $I_{1}=1$ mA, and $\nu_{1}$ stepped from
1-3.5 GHz. At each frequency, the magnetization is initialized to
within 0.1$^{\circ}$ of equilibrium (along $\hat{y}$ and canted
1.4$^{\circ}$ out of plane), and evolved for 100 ns to ensure steady
state. The time-averaged change in voltage $\Delta V_\text{MR}$ due to magnetoresistance is then calculated as
\begin{align}
\Delta V_\text{MR} & =\langle I(t)R(t)-I_{0}R_{0}\rangle\\
& =\left\langle \left(I_{0}+I_{1}\cos(2\pi\nu_{1}t)\right)\left(R_{0}+\Delta R(t)\right)-I_{0}R_{0}\right\rangle \\
& =\langle I(t)\Delta R(t)\rangle,
\end{align}
where $R_{0}$ is the undriven resistance, and
\begin{equation}
\Delta R(t)=R_{0}\delta_{\text{AMR}}m_{x}^{2}
\end{equation}
is the time-dependent resistance change due to precession, with $R_{0}\delta_{AMR}=0.2\,\Omega$.
Figure \ref{fig:macrospyn-parametric} shows $\Delta V_\text{MR}$ calculated
using an integer number of oscillations in the last 10 ns of each
simulation. A large-amplitude parametrically driven peak occurs near
the second harmonic of the ferromagnetic resonance above 3 GHz, with a skew toward higher frequencies similar to the data
in Fig.~2(a) of the main text. The magnitude is also quantitatively
similar, corresponding
to in-plane precession about the equilibrium offset ($m_{z}=0.025$)
amplitude of 30$^{\circ}$ when $\Delta V_\text{MR}=120~\upmu$V, the trajectory of which is shown in inset
(i). The peak is positive because the antidamping spin transfer torque
(pointing away from the inset plot's origin) is, on average, larger when
the angle is maximal, though the phase of the resistance oscillations
lags behind the drive as shown in Fig.~\ref{fig:macrospyn-parametric}(ii). Finally, we note the presence of a much smaller directly-driven oscillation at $\nu_\text{FMR}$, arising from the small equilibrium value $\langle{m_z}\rangle$ and the oscillatory current-induced field $\vec{H}_I$. This feature is visible in Fig.~4(a) of the main text.

\subsubsection{Nanowire stray field along NV axis}\label{Sec:Bstray}

\begin{figure}[h!]
	\centering
	\includegraphics[width=0.7\textwidth]{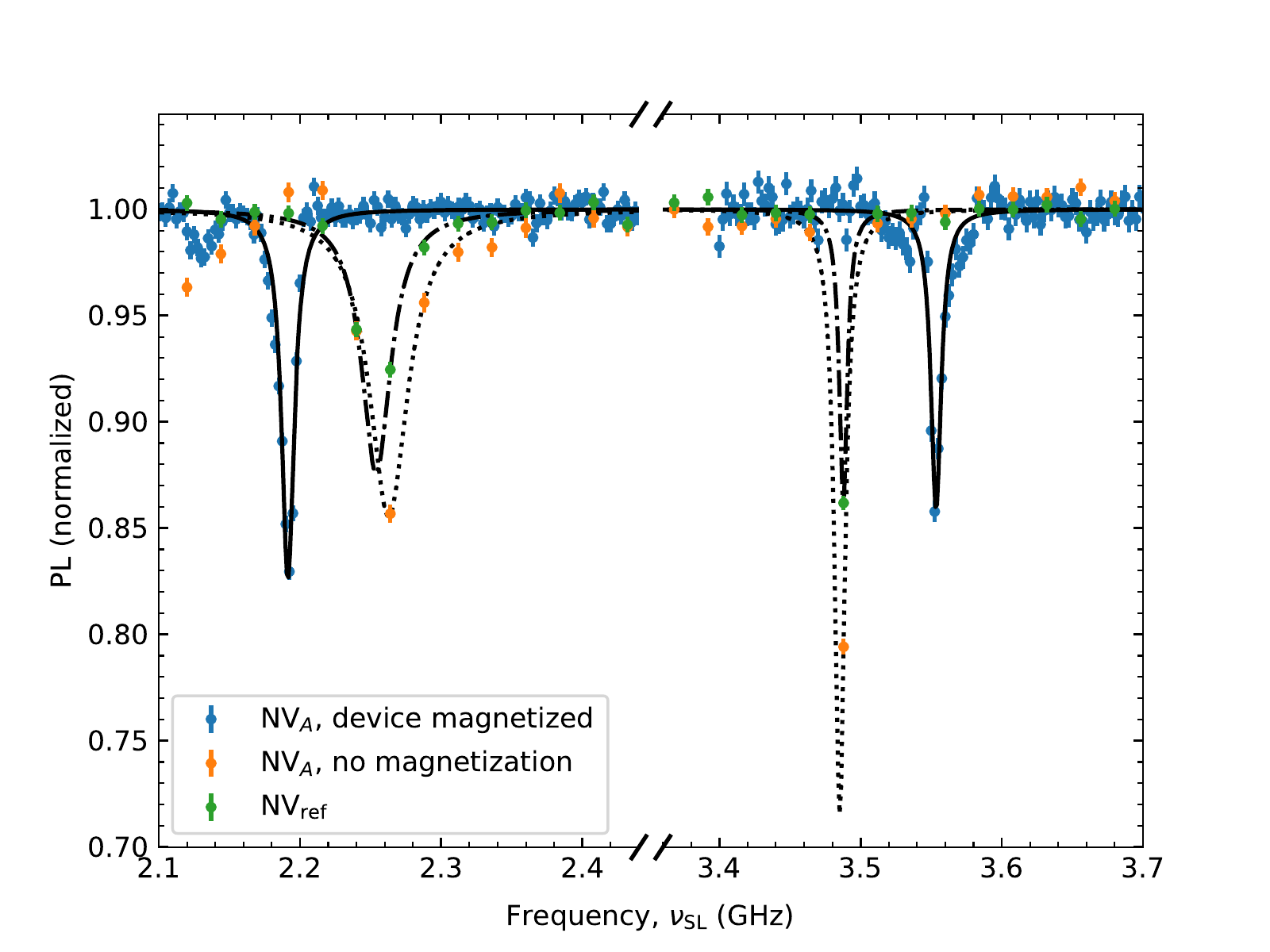}
	\caption{Nanowire stray field as measured by ESR. Measurements of the $m_s = 0 \rightarrow \pm 1$ transitions of NV$_A$ while the device was magnetized, after the device had oxidized, and of NV$_{\text{ref}}$ are fit to Lorentzian profiles in the solid, dotted and dash-dotted lines respectively.}
	\label{Fig:ESRstrayfield}
\end{figure}

In this section, we use the NV's field-dependent electron spin resonance (ESR) to estimate the strength of the stray field along the NV axis when the device is statically magnetized along $\hat{y}$. Figure \ref{Fig:ESRstrayfield} shows the photoluminescence (PL) spectra with field $\mu_0 H_0 = 22.5$~mT applied along the NV$_A$ axis for (blue data) NV$_A$, (green data) a reference NV$_\text{ref}$ having the same orientation as NV$_A$ but positioned 5~$\upmu$m from the device, where the stray field is negligible, and (orange data) NV$_A$ after the nanowire's demise, which removed all evidence of ferromagnetism (we suspect due to oxidization or destruction the Py layer). From the difference in fit ESR frequencies (see Table~\ref{Tab:ESRfits}), we estimate the axial stray field at NV$_A$ most likely lies between $2.1$ and $2.6$ mT, and we use the ``dead device'' value $2.5\pm0.1$ mT as our best estimate.

Note the differences in the linewidths and contrasts are due in part to imperfect power coupling to the stripline, and inhomogeneity in the external field can only account for a maximum of 2~MHz of deviation between NV$_A$ and NV$_\text{ref}$.

\begin{table}[h]
	\centering
	\caption{Fit results from ESR resonances in Fig.~\ref{Fig:ESRstrayfield}. The net NV axial field is estimated assuming a the free electron gyromagnetic ratio $\gamma_{\text{NV}}=2\pi\times 28.0$~GHz/T.}
	\label{Tab:ESRfits}
	\begin{tabular}{l|l|l}
		& $\Delta \nu$ (MHz) & $B_{\parallel}$ (mT)\\ \hline
		NV$_A$, magnetized device & $1362.0 \pm 0.2$ & $24.409 \pm 0.003$ \\
		NV$_A$, oxidized device   & $1222 \pm 4$     & $21.9 \pm 0.1$ \\
		NV$_{\text{ref}}$         & $1230 \pm 10$    & $22.1 \pm 0.2$        
	\end{tabular}
\end{table}

\subsubsection{Estimating the parametric precession angle}

Knowing the strength of the nanowire's stray field at NV$_A$ (Sec.~\ref{Sec:Bstray} above) allows us to estimate the parametric oscillation angle $\Delta \theta$ from the reduction in stray field shown in Fig.~3(a) of the main text, using the macrospin approximation. As discussed above, the steady-state trajectory is roughly sinusoidal and highly confined to the $xy$ plane, so we approximate for simplicity
\begin{align}
m_x & \approx \sin \theta (t),\\
m_y & \approx \cos \theta (t),\\
m_z & \approx 0,\\
\theta (t) & \approx \Delta \theta \cos (\pi \nu_\text{NW} t + \psi), \label{Eq:MagAngleTraj}
\end{align}
where $\theta(t)$ is the time-dependent in-plane angle from $\hat{y}$, $\Delta \theta$ is the steady-state amplitude, and $\nu_{\text{NW}}$ nanowire's drive frequency, equal to twice the parametric response frequency, and $\psi$ is a steady-state phase shift. In this limit, we can calculate the time-averaged magnetization along $\hat{y}$, which will reduce the stray field experienced by NV$_A$ as
\begin{align}
\langle m_{y}\rangle&\approx\left\langle \cos\left(\Delta\theta\cos(\pi\nu_{\text{NW}}t+\psi)\right)\right\rangle \\
&=J_0(\Delta \theta) \label{Eq:Mavg_approx}
\end{align}
where $J_0(\Delta \theta)$ ($\approx 1-\frac{1}{4}\Delta\theta^2$) is the zeroth order Bessel function. Figure 3(a) in the main text shows an increase in the $m_s=0 \rightarrow -1$ transition frequency of $\Delta \nu_- = 13$~MHz, which corresponds to a decrease in stray field of $\Delta B_{\text{stray}} \approx 0.5$~mT. Assuming the precession is sufficiently symmetric that the stray field orientation remains the same, the fractional change
\begin{align}
\frac{\Delta B_{\text{stray}}}{B_{\text{stray}}}&\approx1-\langle m_{y}\rangle.
\end{align}
Using $B_\text{stray}=2.5$~mT from Sec.~\ref{Sec:Bstray} at the same external field, we estimate $\Delta \theta\approx 60^\circ$. 

Under the same approximations, we can independently estimate $\Delta \theta$ from the magnetoresistance signal
\begin{align}
\Delta R(t) = R_0\delta_{\mathrm{AMR}} \sin^2 \left( \Delta \theta \cos (\omega_r t + \varphi) \right).
\end{align}
When current $I(t) = I_0 + I_{\mathrm{RF}} \cos( \omega t )$ is sent through the device, with $I_0\gtrsim 4$~mA and $I_\text{RF}\sim 1$, time-averaging the precession-induced voltage yields
\begin{align}
\Delta V_\text{MR} &\approx  \frac{1}{2}R_{0}\delta_{\text{AMR}}I_{0}\left(1-J_{0}(2\Delta\theta)\right),
\end{align}
where we have dropped the comparatively small ``mixdown'' term involving $I_\text{RF}$ for simplicity (the $I_\text{RF}$ term contributes $\lesssim 10\%$ to the total signal for this range of parameters, as shown in supplementary Sec.~\ref{subsec:macrospin-parametric}). In this limit, the peak measurement of $\Delta V_\text{MR} = 360$~$\upmu$V in Fig.~3(a) of the main text ($I_0=4.9$~mA, $I_{1}= 1.15$~mA, $R_0 \delta_\text{AMR}= 0.2~\mathrm{\Omega}$) corresponds to $\Delta \theta \approx 55^\circ$. Similarly, the $120~\upmu$V parametric peak in Fig.~2(a) of the main text ($I_0=4.08$~mA) corresponds to $\Delta\theta\approx 32^\circ$, in reasonable agreement with the macrospin simulation (Sec.~\ref{subsec:macrospin-parametric}).

\pagebreak
\begin{figure}[h!]
	\centering
	\includegraphics[width=14cm]{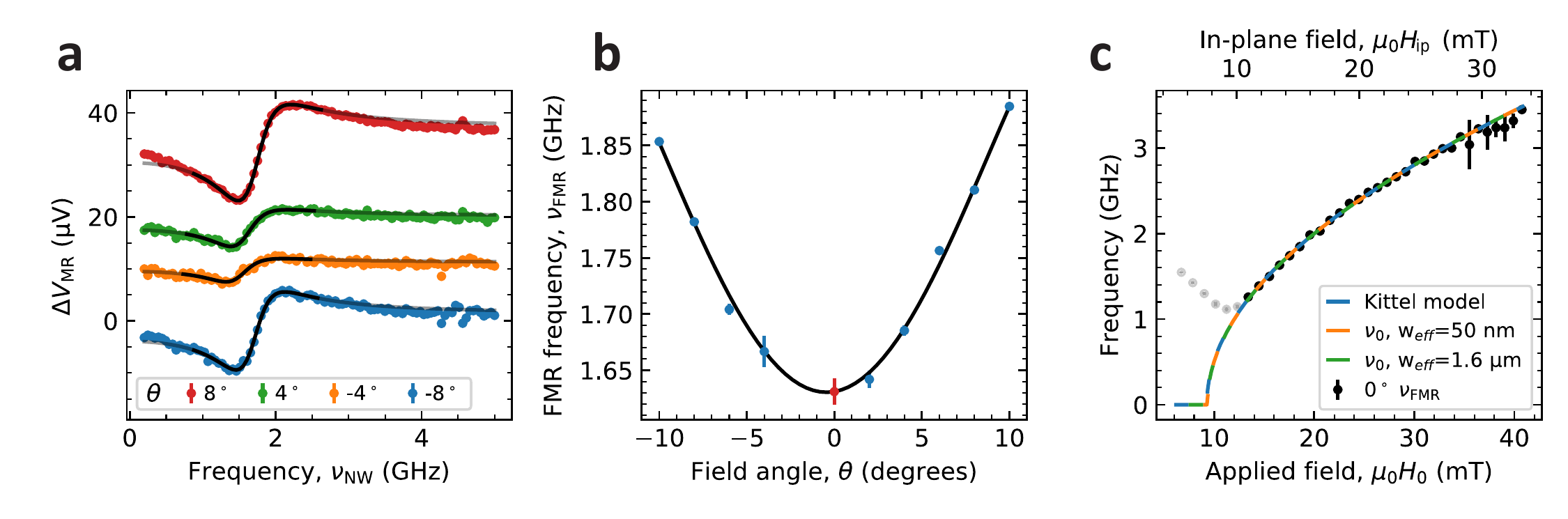}
	\caption{Estimating the ferromagnetic resonance (FMR) frequency $\nu_\text{FMR}$ at in-plane angle $\theta=0$. (a) Spin-transfer-driven FMR spectra for $\theta=\pm 4^\circ$ and $\pm 8^\circ$ for static field $\mu_0 H_0 = 16.5$~mT applied $35^\circ$ out of plane, as in the main text. Traces are offset for clarity. Dark lines show Fano fits, and gray lines extend these fits outside the chosen range of frequencies. (b) Fitted frequencies (blue points) over the full range of angles, fit to a symmetric quartic function. (c) Summary of so-estimated $\nu_\text{FMR}(0)$ $0^\circ$, with the range 14-35~mT fit with a simple Kittel model (blue curve), as well as a lowest-order spinwave model assuming the ``extreme'' values of fixed effective widths 50~nm (orange curve) and 1.6~$\upmu$m (green curve).}
	\label{Fig:FMRfits}
\end{figure}
\section{Ferromagnetic resonance frequencies}\label{Sec:FMRfit}
Because the spin transfer drive and magnetoresistance signal both approximately vanish when the equilibrium value of $\hat{m}$ is (nearly) parallel to $\hat{y}$ (i.e., the field's in-plane angle $\theta=0$, we cannot measure the ferromagnetic resonance (FMR) frequency $\nu_\text{FMR}$ directly. Instead, we measure $\nu_\text{FMR}(\theta)$ for a set of 8 to 9 small angles spanning $\pm 10^\circ$ (maintaining the out-of-plane angle $35^\circ$ as in the main text), and then fit the resulting frequencies to a symmetric polynomial to estimate $\nu_\text{FMR}(0)$ (and a misalignment angle $\theta_0$).

Figure \ref{Fig:FMRfits}(a) shows a ``typical'' set of spin-transfer-driven FMR spectra with applied field $\mu_0 H_0=16.5$~mT, taken as discussed in Sec.~\ref{subsec:Single-shot-lock-in-measurement}, with the Joule heating background (Sec.~\ref{sec:joule-background}) subtracted. Due to the frequency-dependence of the drive current (see Sec.~\ref{subsec:Estimating-RF-current}), we fit only the data near the resonant feature to Eq.~\ref{eq:macrospin-FMR-fit-function} to extract the frequency $\nu_\text{FMR}$ and width $\Delta\nu$. As expected, the signal increases with $|\theta|$. Also, as shown in Fig.~\ref{Fig:FMRfits}(b), the frequency decreases as $\theta$ approaches zero, consistent with the shape anisotropy maximally opposing the applied field at $\theta = 0$. Exploiting the mirror symmetry of our geometry, we fit the observed angular dependence in Fig.~\ref{Fig:FMRfits}(b) to a low-order symmetric polynomial of the form
\begin{equation}
\nu_\text{FMR}(\theta) = C_0 + C_2 (\theta - \theta_0)^2 + C_4 (\theta - \theta_0)^4,
\end{equation}
with fit constants $C_0$, $C_2$, $C_4$, and offset angle $\theta_0$. The offset angle ($\theta_0 = 0.4^\circ \pm 0.1^\circ$ for the shown data set) takes on values within $\pm0.5^\circ$ of $0^\circ$ over the usable range of applied fields (14-35~mT). % Note the signal-to-noise makes more of the small-angle data unreliable at high applied field, and so the -4$^\circ$ and +2$^\circ$ data was ignored for $\mu_0 H_0 > 22.5$.
Completing the same analysis at each value of applied field produces the zero-angle frequency data $\nu_\text{FMR}(0)$ shown in  Fig.~\ref{Fig:FMRfits}(c). Clear fit systematics preclude the trustworthiness of frequencies so estimated below 14~mT. The ``reliable'' region (dark symbols) can then be fit to a variety of models to estimate material parameters of the permalloy (Py) layer. 

\subsection{Macrospin approximation to resonant frequency}\label{subsec:macrospin-approximation}

To gain immediate insight, we first assume the magnetization simply behaves as a uniformly-magnetized ellipsoid with equilibrium magnetization $\hat{m}$ approximately parallel to $\hat{y}$. As derived in Sec.~\ref{subsec:macrospin-small-angle-precession}, the resonant frequency for our geometry (Eq.~\ref{eq:macrospin-FMR-frequency})
\begin{equation}
\nu_\text{FMR} = \frac{\gamma_0 \mu_0}{2\pi} \sqrt{(H_y - H_{yx})(H_y + H_{zx})},
\end{equation}
where $H_y = H_0 \cos(35^\circ)$ is the in-plane component of the applied field. Fitting the ``reliable'' range of data (blue curve in Fig.~\ref{Fig:FMRfits}(c)) yields effective saturation fields $\mu_0 H_{yx}= 7.57\pm0.08$~mT and $\mu_0 H_{zx}= 517\pm4$~mT. The low value of $H_{zx}$ (nominally close to the saturation magnetization) suggests non-uniform magnetic dynamics and / or other nonidealities of the Py layer. 

\subsection{Lowest-order spinwave approximation to resonant frequency}\label{subsec:lowest-spinwave}

To get a sense of scale for the potential impact of nonuniform dynamics, we can also perform a fit to an approximate lowest-order (most uniform) spinwave resonance, which has frequency \cite{Kalinikos1986Theory, Bracher2017Parallel}
\begin{equation}\label{Eq:WaveDispersion}
\nu_{\mathbf{k}} = \frac{\gamma_0 \mu_0}{2\pi} \sqrt{\left( H_y + M_\text{s} \lambda_\text{ex} k^2 - H_\text{d} \right) \left( H_y + M_\text{s} \lambda_\text{ex} k^2 - H_\text{d} + M_\text{s} F_{\mathbf{k}} \right)},
\end{equation}
where $H_y = H_0 \cos(35^\circ)$ is the in-plane component of the applied field, $H_\text{d}$ is an effective demagnetizing (dipole) field, $M_\text{s}$ is Py's saturation magnetization, $\lambda_\text{ex}=2A_\text{ex}/(\mu_0 M_\text{s}^2)$, with exchange constant $A_\text{ex}=1.05 \times 10^{-11}$ J/m \cite{Smith1989Magnetoresistive}, $k$ is the magnitude of the spin wave vector $\mathbf{k}=k_x \hat{x}+ k_y \hat{y}+ k_z \hat{z}$, and  
\begin{equation}\label{Eq:Fk}
F_{\mathbf{k}} = 1 + g_k \left( \sin^2 \theta_k - 1 \right) + \frac{M_\text{s} g_k (1 - g_k) \sin^2 \theta_k}{H_y - H_\text{d} + M_\text{s} \lambda_{ex} k^2},
\end{equation}
where $\theta_k$ is the angle between the equilibrium magnetization orientation $\langle\hat{m}\rangle$ and $\mathbf{k}$, and $g_k = 1 - \left(1 - e^{- k t_\text{Py}} \right) /(k t_\text{Py})$ with device thickness $t_\text{Py}$. 

For our thin film, we expect $\hat{m}$ to be approximately uniform along $z$ \cite{Kalinikos1986Theory, Bayer2005Spin} (we also ignore the small offset in the equilibrium out-of-plane component $\langle m_z\rangle$). The nanowire further constrains the longitudinal wavenumber as $k_x = n_x \pi / L_\text{eff}$, with $n_x = 1,2,3, \dots$ and $L_\text{eff} \approx 8.05 ~\upmu$m deviating slightly from the geometrical length due to effective dipolar boundary conditions on the wire \cite{Bayer2005Spin, Guslienko2002Effective}. The remaining relevant wave number is often written $k_y = n_y \pi/w_\text{eff}$ in terms of the transverse mode number $n_y = 1,2,3, \dots$ and an effective width $w_\text{eff}$, which we expect to be \emph{smaller} than the actual wire width when $\hat{m} \parallel \hat{y}$ \cite{Bracher2017Parallel, Bayer2005Spin, Bracher2017Creation}. The lowest frequency (and most spatially homogeneous) mode should have $n_x = n_y = 1$, so we treat the resonances measured in transport as having a wave vector $\mathbf{k}_\text{FMR} = \frac{\pi}{L_\text{eff}} \hat{x}+ \frac{\pi}{w_\text{eff}} \hat{y}$.

Equation \ref{Eq:WaveDispersion} effectively contains two fit parameters, $M_\text{s} \lambda_\text{ex} k^2 - H_\text{d}$ and $M_\text{s} F_\mathbf{k}$, which themselves are composed of three unknown quantities, $w_\text{eff}$, $M_\text{s}$, and $H_\text{d}$. Table \ref{Tab:magnonfits} shows the fit values of $H_\text{d}$ for a wide range of assumed $w_\text{eff}$, with the corresponding fits for $w_\text{eff} = 50$ nm (orange) and $w_\text{eff} = 1.6 ~\upmu$m (green) plotted in Fig.~\ref{Fig:FMRfits}(c) showing negligible deviation from the macrospin model. As expected, as $w_\text{eff}$ becomes large, the $H_\text{d}\rightarrow H_{yx}$ and $M_\text{s}\rightarrow H_{yx}+H_{zx}$ from the macrospin approximation (Sec.~\ref{subsec:macrospin-approximation}). We presume $w_\text{eff}$ should not be smaller than 50~nm, and the low value of $\mu_0 M_\text{s}$ still suggests some nonidealities in the Py layer, which will be the subject of future investigations and higher quality materials deposition. Note that modifying the parameters $M_\text{s}$, $H_\text{d}$, $w_\text{eff}$ or $A_\text{ex}$ by factors of order unity does not affect the key results -- observed spin transfer threshold for parametric oscillations, observed stray fields (or lack thereof) at the NV, and observed spin transfer damping -- of the main text. 

\begin{table}[h]
	\centering
	\caption{Fit results of lowest-ordered magnon modes with fixed $w_\text{eff}$ to data in Fig.~\ref{Fig:FMRfits}(c).}
	\label{Tab:magnonfits}
	\begin{tabular}{l|l|l}
		Fixed $w_\text{eff}$ (nm)  & $\mu_0 H_\text{d}$ (mT) & $\mu_0 M_\text{s}$ (mT)\\ \hline
		50 & $155 \pm 1$ &  $709 \pm 6$ \\
		100 & $50.1 \pm 0.3$ & $613 \pm 5$ \\
		200 & $19.17 \pm 0.04$ &  $569 \pm 5$ \\
		400 & $10.81 \pm 0.06$ & $548 \pm 5$ \\
		800 & $8.79 \pm 0.08$ & $537 \pm 5$ \\
		1600 & $8.58 \pm 0.09$ & $532 \pm 5$ \\       
	\end{tabular}
\end{table}

\pagebreak
\section{Magnon-induced NV spin relaxation rates} \label{Sec:RelaxationRates}

\begin{figure}[h!]
	\centering
	\includegraphics[width=12cm]{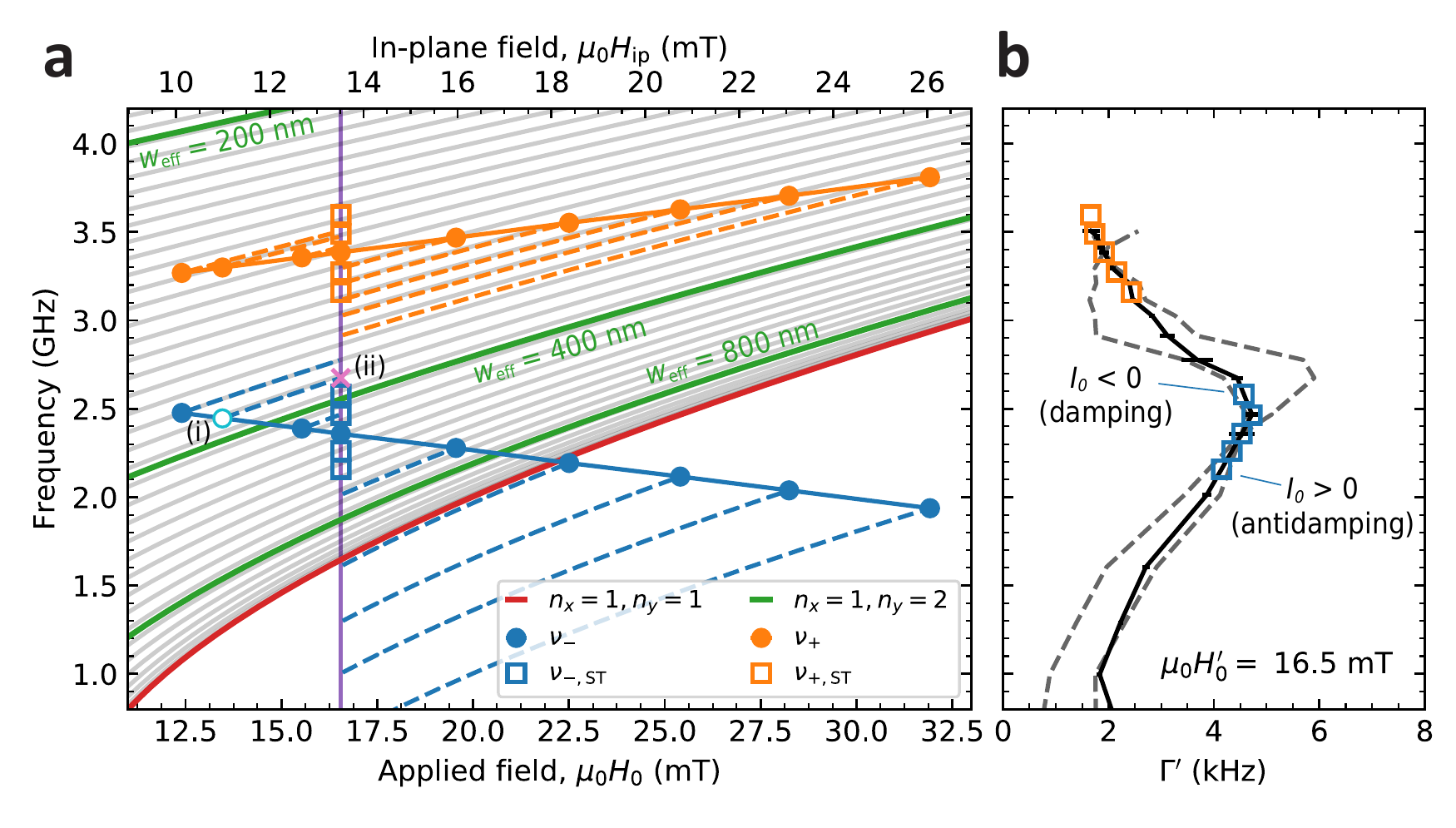}
	\caption{Reconstructing the field and frequency dependence of the NV spin relaxation rate $\Gamma^\prime$ in the absence of spin transfer effects. 
		(a) Spectrum of spin wave mode frequencies for the first transverse mode ($n_y=1$, $n_x=1,2,3,...$, gray lines), and NV resonance frequencies $\nu_+$ (orange) and $\nu_-$ (blue) when varying only the field (filled circles) or varying the DC bias while compensating the field at the Py layer (hollow squares). The fundamental mode  ($n_x=n_y=1$) is highlighted in red, and the green lines correspond to the second fundamental transverse mode ($n_x=1, n_y=2$) for different assumed effective widths $w_\text{eff}$ (labeled); all modes follow the same family of curves to good approximation (deviating from each other by much less than a spin wave linewidth) over the studied range. The vertical line at $\mu_0 H_0= 16.5$~mT indicates the field at which the spin transfer effects were probed in Fig.~5(right) of the main text. Dashed lines highlight which NV measurements (filled symbols) are used to estimate $\Gamma^\prime$ at which frequencies along the vertical line. 
		(b) Reconstructed spin-transfer-free relaxation rates (solid line) along the vertical line cut in (a) with error bars from the bias-free measured values (solid points in (a)). The dashed lines represent the absolute (and quite extreme) bounds of the analysis. Orange (blue) squares correspond to $\Gamma^\prime_+$ ($\Gamma^\prime_-$) in Fig.~5(b) of the main text.}
	\label{Fig:GammaReconstruction}
\end{figure}

In this section, we use the spin-transfer-free relaxation rates $\Gamma(\nu,H)$ measured at a variety of NV probe frequencies $\nu$ and applied fields $H$ to estimate the rates at other values of $\nu$ and $H$ (i.e., the color scales in Fig.~5 of the main text). The basic idea can be understood by inspecting the phase space of spin wave modes shown in Fig.~\ref{Fig:GammaReconstruction}(a). Here, many spin wave mode frequencies $\nu_\textbf{k}$ (where $\mathbf{k}=k_x \hat{x}+k_y\hat{y}+k_z\hat{z}$ is the mode wavenumber; see Sec.~\ref{subsec:lowest-spinwave}) are plotted over our range of applied fields. Importantly, all modes follow the same family of curves to good approximation, meaning one can use a measurement of $\Gamma(\nu,H)$ (taken at the filled symbols) to estimate the relaxation rate $\Gamma^\prime(\nu^\prime,H^\prime)$ at another location along the nearest $\nu_\mathbf{k}$ curve; the quantities that vary the most along these curves are the field, frequency, density of states, and thermal occupancy, all of which are known or can be approximated, as discussed below.

\begin{figure}[h!]
	\centering
	\includegraphics[width=10cm]{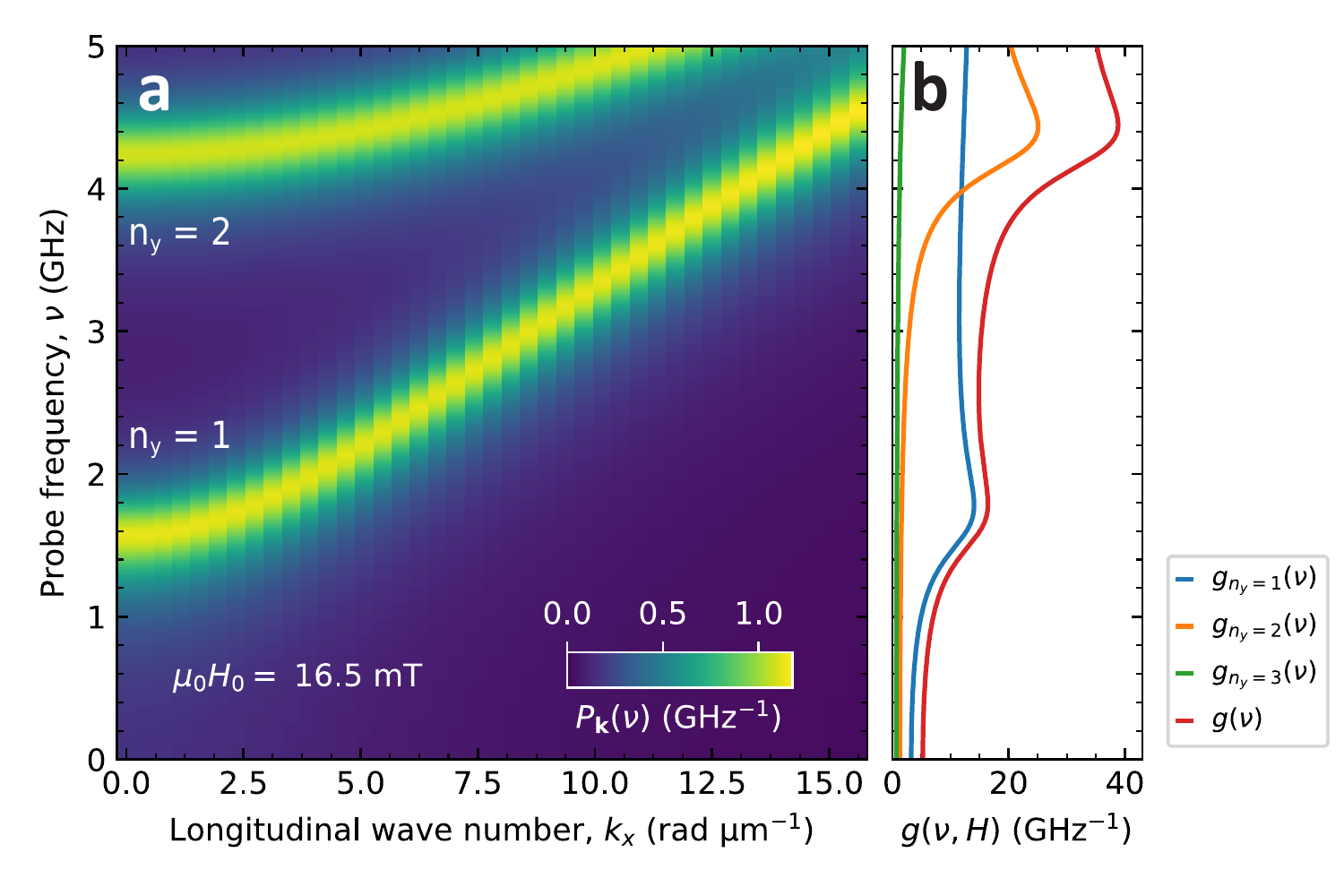}
	\caption{Spin wave dispersion and broadened density of modes assuming an effective width $w_\text{eff}=200$ nm. (a) Noise distribution $P_\mathbf{k}$ versus frequency $\nu$ and longitudinal wavenumber $k_x$, showing the first $(n_y=1)$ and second $(n_y=2)$ transverse modes at an applied field of $\mu_0 H_0=16.5$~mT along the NV axis. (b) Broadened density of modes $g(\nu)$, and the contributions from the first ($g_{n_y=1}(\nu)$), second ($g_{n_y=2}(\nu)$), and third ($g_{n_y=3}(\nu)$) modes.}
	\label{Fig:MagnonDOS}
\end{figure}

First, we note that the \emph{total} NV spin relaxation rate \cite{Slichter1990Principles}
\begin{align}
\Gamma (\nu, H) &= \Gamma^0 + \frac{\gamma_\text{NV}^2}{2} S_\perp (\nu, H)\\
& \approx \frac{\gamma_\text{NV}^2}{2} S_\perp (\nu, H)
\end{align}
comprises the sum of the NV's (small) internal rate $\Gamma^0_{-(+)} = 64 \pm 7$~Hz $(54 \pm 12$~Hz) (as measured with NV$_A$ at $\mu_0 H_0 = 22.5$~mT after the device magnetization disappeared) and the rate $\gamma_\text{NV}^2 S_\perp/2~(\sim $ kHz) driven by the magnetization's stray field noise power spectral density $S_\perp(\nu,H)$ (units of T$^2$/Hz), where $\gamma_\text{NV}$ is the magnitude of the NV spin's gyromagnetic ratio and the subscript $\perp$ reminds us that it is the fields perpendicular to the NV axis that drive the transitions. Each spin wave mode contributes noise power in proportion to its thermal occupancy $\bar{n}(\nu_\mathbf{k})$, and so we can write
\begin{equation}\label{eq:Sperp1}
S_\perp (\nu,H) = \sum_\mathbf{k} \bar{n}(\nu_\mathbf{k}) f_\mathbf{k}P_\mathbf{k}(\nu,H),
\end{equation}
where $f_\mathbf{k}$ is a mode-dependent geometrical constant converting occupancy to noise power at the NV (units of T$^2$/magnon), and $P_\mathbf{k}(\nu,H)$ is a unity-normalized density function (units of 1/Hz) describing how this power is distributed over the frequency domain (i.e., a normalized version of the mode's power susceptibility, such as Eq.~\ref{eq:brownian-noise-spectrum}).

To simplify the analysis, we assume the mode profiles do not change much over our range of applied fields, so that $f_\mathbf{k}$ is approximately independent of field. We \emph{do} expect $f_\mathbf{k}$ to depend on $\mathbf{k}$ and the exact location of the NV relative to the nanowire, taking on the largest values when $1/|\mathbf{k}|$ is comparable to the wire-NV distance $d$ \cite{Van2015Nanometre, Du2017Control}. 

To simplify further, we note that the observed linewidth $\Delta\nu\sim600$~MHz of the fundamental mode is approximately constant over our field range, consistent with the behavior predicted by the macrospin approximation (Sec.~\ref{subsec:macrospin-linear-susceptibility}). We therefore assume the distributions $P_\mathbf{k}(\nu,H)$ depend only on $\nu_\mathbf{k}(H)$ and the probe frequency $\nu$. Figure~\ref{Fig:MagnonDOS}(a) shows an example $P_\mathbf{k}$ for applied field $\mu_0 H_0=16.5$~mT along the NV axis. Summing across $k_x$, we then define a total ``linewidth-broadened density of modes'' $g(\nu,H)$, which is shown in Fig.~\ref{Fig:MagnonDOS}(b). 

\begin{figure}[h!]
	\centering
	\includegraphics[width=12cm]{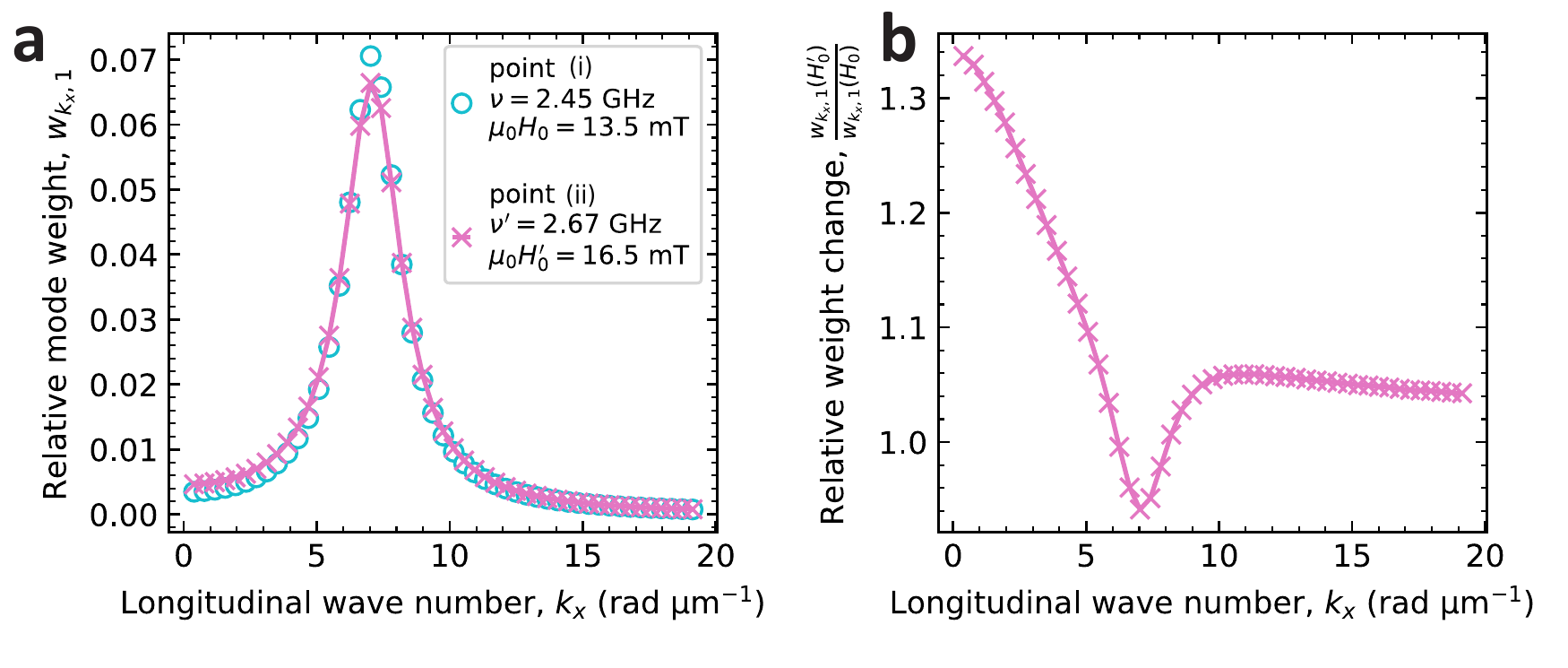}
	\caption{Magnon transfer function weights at points (i) and (ii) in Fig.~\ref{Fig:GammaReconstruction}(a). 
		(a) Relative magnon weights $w_{\mathbf{k}}$ probed at frequencies $\nu_{\mathbf{k}}=2.45$~GHz (13.5 mT) and $\nu_{\mathbf{k}}^\prime=2.67$~GHz (16.5 mT) versus $k_x$, which peaks at $k_x=7$~rad/$\upmu$m (as indicated in Fig.~\ref{Fig:MagnonDOS} for $\mu_0 H_0^\prime = 16.5$~mT).
		(b) Ratio of weights for these two fields. 
		}
	\label{Fig:MagnonTransferWeights}
\end{figure}

If we imagine following one of the $\nu_\mathbf{k}(H)$ curves in Fig.~\ref{Fig:GammaReconstruction}(a), we notice two important quantities change -- the mode frequencies $\nu_\mathbf{k}(H)$ and the density of states $g(\nu, H)$, suggesting that, if we factor  these trends from  Eq.~\ref{eq:Sperp1}, we can write
\begin{equation}\label{eq:Sperp2}
S_\perp (\nu,H) = \bar{n}(\nu) g(\nu,H) \sum_\mathbf{k}  w_\mathbf{k}(\nu, H) f_\mathbf{k},
\end{equation}
in terms of a ``weighting factor''
\begin{align}
w_\mathbf{k}(\nu, H) \equiv \frac{\bar{n}(\nu_\mathbf{k}) P_\mathbf{k}(\nu,H)}{\bar{n}(\nu)g(\nu,H)}
\end{align}
that should be fairly insensitive to the distance traveled along a given $\nu_\mathbf{k}$ curve. Of particular relevance to the cooling argument of the main text, Fig.~\ref{Fig:MagnonTransferWeights}(a) shows these weights for the direct measurement at point (i) in Fig.~\ref{Fig:GammaReconstruction}(a) (cyan circles) as well as the location of maximum cooling (ii) (magenta markers).  As we have engineered, both peaks occur at the same value of $k_x$ (7~rad/$\upmu$m), which is a restatement of the fact that we have moved along a $\nu_\mathbf{k}$ curve. Importantly, the distributions at these two extremes look very similar, with individual weights differing by at most 34\%, as shown in Fig.~\ref{Fig:MagnonTransferWeights}(b). We remind ourselves that these weights are multiplied by values of $f_\mathbf{k}$ expected to oscillate with $k_x$ underneath a smooth envelope, such that the summed effect is more likely of order the $w_\mathbf{k}$-weighted average of the ratio in Fig.~\ref{Fig:MagnonTransferWeights}(b) (i.e., $\sim$3\%). The presence of additional transverse modes having comparable values of will \emph{not} have an impact beyond that of Fig.~\ref{Fig:MagnonTransferWeights}(b) unless their mode frequencies deviate from the fundamental mode's family of $\nu_\mathbf{k}(H)$ curves (gray lines in Fig.~\ref{Fig:GammaReconstruction}(a)) by an amount comparable to $\Delta \nu$ over the length of the dotted blue line. Even including a more complicated spin wave model or micromagnetic simulations, we do not expect to find modes so strongly deviating from these trends over so small a field range.

We can now convert the directly-measured values of $\Gamma$ at frequencies $\nu$ and fields $H$ (filled circles in Fig.~\ref{Fig:GammaReconstruction}(a)) into estimates at other values $\nu^\prime$ and $H^\prime$ using the ratio
\begin{equation}\label{eq:gamma-ratio}
\frac{\Gamma^\prime(\nu^\prime,H^\prime)}    
{\Gamma(\nu,       H)} = 
\frac{S_\perp(\nu^\prime,H^\prime)}    
{S_\perp(\nu,       H)} =
\frac{\bar{n}(\nu^\prime)}{\bar{n}(\nu)}
\frac{g(\nu^\prime,H^\prime)}{g(\nu,H)}
\frac{\sum_\mathbf{k}  w_\mathbf{k}(\nu^\prime, H^\prime) f_\mathbf{k}}{\sum_\mathbf{k}  w_\mathbf{k}(\nu, H) f_\mathbf{k}}.
\end{equation}
The first ratio $\bar{n}(\nu^\prime)/\bar{n}(\nu)=(e^{h\nu/k_B T}-1)/(e^{h\nu^\prime/k_BT}-1)$, with Planck constant $h$, Boltzmann constant $k_B $, and temperature $T$. The second ratio can be calculated as in Fig.~\ref{Fig:MagnonDOS}, and the final ratio should be comparable to 1, as discussed above. The resulting reconstruction is plotted (solid line) for $\mu_0 H_0 = 16.5$~mT, i.e., the field at which the spin transfer measurements were made in Fig.~\ref{Fig:GammaReconstruction}(b). Similar curves can be generated at other values of field, as is plotted on the color scales in Fig.~5 of the main text. Also plotted in Fig.~\ref{Fig:GammaReconstruction}(b) are upper and lower bounds (dashed curves) corresponding to worst-case-scenario systematic errors in the final ratio, calculated by assuming the \emph{only} contributing mode is the one having the \emph{largest} and \emph{smallest} values of $w_\mathbf{k}(H^\prime)/w_\mathbf{k}(H)$ in Fig.~\ref{Fig:MagnonTransferWeights}(b). Importantly, the expected bias-dependent changes in $\Gamma^\prime_\pm$ for Fig.~5(b) of the main text (the probe frequencies of which are indicated by orange and blue squares in Fig.~\ref{Fig:MagnonTransferWeights}) is much smaller than what is observed, and has the opposite trend with applied current.

\begin{figure}[h!]
	\centering
	\includegraphics[width=10cm]{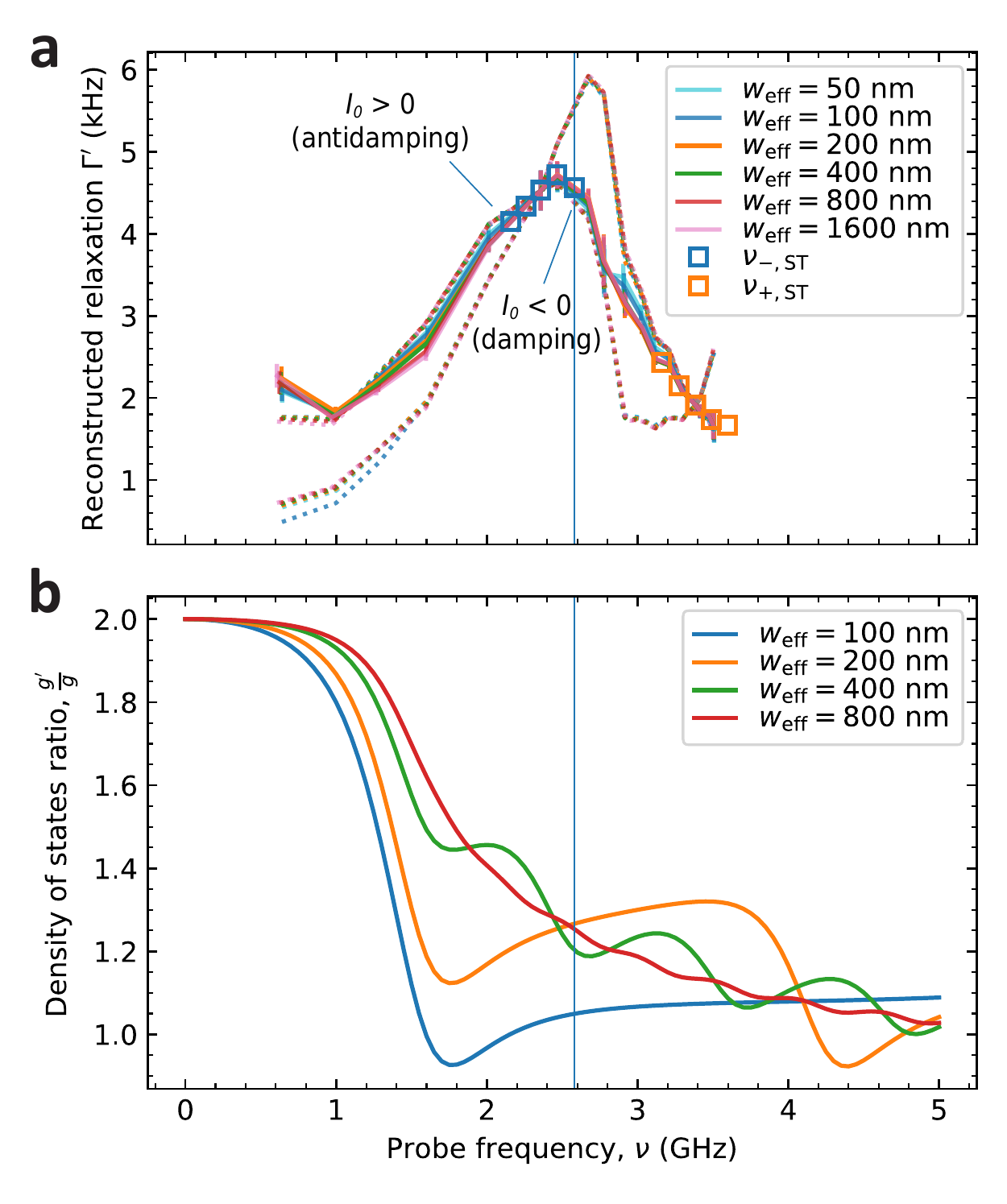}
	\caption{Reconstructed relaxation rates $\Gamma^\prime$ for a range of effective widths $w_\text{eff}$. (a) $\Gamma^\prime$ at $\mu_0 H_0 = 16.5$~mT for different assumed values of $w_\text{eff}$, illustrating a minimal impact from changes in mode structure, in particular near the spin-transfer probe frequencies of $\nu_{\pm, \text{ST}}$.
		(b) Ratio of the linewidth-broadened density of states at the same field for a factor of 2 increase in $\Delta \nu$, illustrating that the total density of states increases at the probe frequency $\nu_- = 2.58$~GHz of maximal cooling (vertical line).} 
	\label{Fig:Reconstruction_weffs}
\end{figure}

Figure \ref{Fig:Reconstruction_weffs}(a) shows the same calculations for effective widths spanning a wide range. As expected, the presence of different transverse mode structure has little effect on these estimates. 

Note that, below the fundamental resonance, there is no obvious choice for the blue dashed curves, but there \emph{are} still modes whose tails contribute to $S_\perp$. As such, we have chosen to fix the \emph{difference} from the fundamental mode frequency $\nu_\text{FMR}$ (where the density of modes is highest). In this region, the approximation that the final ratio in Eq.~\ref{eq:gamma-ratio} is $\approx 1$ becomes increasingly incorrect -- mainly because the occupation at the probing frequency no longer matches the occupation of the nearest magnon modes -- as evidenced by $\Gamma^\prime$ exceeding the upper bounds at frequencies below $\nu_\text{FMR} = 1.64$~GHz.

This ``spin-transfer-free'' approximation is valid to the specified tolerances provided the linewidth $\Delta\nu$ is constant. When damped by spin transfer, however, we expect the linewidth to broaden, which can redistribute noise away from $\nu_\mathbf{k}$ and potentially reduce $\Gamma$ without cooling. However, as shown in Fig.~\ref{Fig:Reconstruction_weffs}(b), the broadened density of modes actually increases under the conditions of maximal spin transfer cooling (vertical line), where the linewidth changes by at most a factor of $\sim$2. In the worst shown scenario, where the effective width $w_\text{eff}= 100$ nm, such that the second transverse mode is well above the probe frequency, the broadened density of modes at the probe frequency $\nu_-=2.58$~GHz still increases due to the tails of the modes away from $\nu_-$. As $w_\text{eff}$ increases, the higher-order transverse mode frequencies approach $\nu_-$, and $g$ is found to increase by as much as $\sim25\%$. Therefore, if we assume that $f_\mathbf{k}$ does not vary significantly over the resonance linewidth (and / or varies linearly), then the sum in Eq.~\ref{eq:Sperp2} should remain roughly the same or increase (in opposition to the observed trend) as a result of the increased $\Delta\nu$. Combined with the fact that the maximum temperature change due to Joule heating $\Delta T\lesssim 5$ K (see Sec.~\ref{subsec:joule-maximum-temperature}), we expect this is not a dominant issue.

\bibliographystyle{bibstyle-jack}
\bibliography{errthing}